\documentclass[superscriptaddress,letterpaper]{revtex4}
\usepackage{graphicx}
\usepackage{graphics}

\def\mgut{M_{\rm GUT}}
\def\br{{\rm BR}}
\def\msusy{M_{\rm Susy}}

\def\be{\begin{equation}}
\def\ee{\end{equation}}
\def\bea{\begin{eqnarray}}
\def\eea{\end{eqnarray}}

\def\BR{{\rm BR}}
\def\ee{\mbox{$e^+e^-$}}
\def\bb{\mbox{$b\bar{b}$}}

\def\hsm{h_{\rm SM}}
\def\hl{h}
\def\hh{H}
\def\ha{A}
\def\tanb{\tan\beta}

\def\epem{e^+e^-}
\def\gev{~{\rm GeV}}
\def\tev{~{\rm TeV}}
\def\fbi{~{\rm fb}^{-1}}

\def\rts{\sqrt s}

\def\mhsm{m_{\hsm}}
\def\mhh{m_{\hh}}
\def\mha{m_{\ha}}
\def\mhl{m_{\hl}}

\def\anti{\overline}
\def\gtot{\Gamma_{\rm tot}}

\def\lsim{\mathrel{\raise.3ex\hbox{$<$\kern-0.75em\lower1.1ex\hbox{$\sim$}}}}
\def\gsim{\mathrel{\raise.3ex\hbox{$>$\kern-0.75em\lower1.1ex\hbox{$\sim$}}}}

\begin{document}

\bibliographystyle{revtex}

\title{The Precision of Higgs Boson Measurements and Their Implications }

\author{{\bf The Precision Higgs Working Group of Snowmass 2001\\}
J. Conway}
\affiliation{Dept. of Physics, Rutgers University, Piscataway, NJ USA}
\author{K. Desch}
\affiliation{Dept. of Physics, Univ. of Hamburg}
\author{J.F. Gunion}
\affiliation{Davis Institute for HEP, Univ. of California, Davis, CA  USA}
\author{S. Mrenna}
\affiliation{Fermilab, Batavia, IL  USA}
\author{D. Zeppenfeld}
\affiliation{Department of Physics, U. of Wisconsin, Madison, WI  USA}


\date{\today}

\begin{abstract}
The prospects for a precise exploration of the properties of 
a single or many observed Higgs bosons at future accelerators 
are summarized, with particular emphasis on the abilities of 
a Linear Collider (LC).  Some implications of these measurements
for discerning new physics beyond the Standard Model (SM) are also
discussed.
\end{abstract}

\maketitle

\section{INTRODUCTION}

Despite the experimental verification of the
Electroweak Symmetry-Breaking pattern $SU(2)_L\times U(1)_Y\to U(1)_{em}$,
the origin of the Higgs mechanism remains unknown.  The simplest
proposal is the existence of a fundamental, complex scalar field which
is an $SU(2)_L$ doublet, the neutral component of which
acquires a vacuum expectation value $v=246$ GeV.  
The one physical degree of freedom is the Standard Model (SM) 
Higgs boson $\hsm$, which couples to
each fermion and electroweak gauge boson in proportion to its mass.
The analysis of precision Electroweak data strongly suggests the existence of
a light, SM-like Higgs boson, which should be discovered
in the next generation of hadron-collider experiments.  This report
addresses the question of how well the properties of this Higgs boson
can be measured at planned and proposed colliders, and whether 
there are associated experimental signatures
that would reveal additional structure to the Higgs sector.

\section{THE STANDARD MODEL HIGGS BOSON}

The SM Higgs boson is a spin--0, CP-even
scalar with tree level couplings to fermions $\propto m_f/v$
and to the $V=W$ and $Z$ bosons $\propto M_V\times M_W/v$.  The
Higgs boson mass $\mhsm$ is
$v\sqrt{\lambda}$, where $\lambda$ is the self-coupling of the
Higgs field that also sets the strength of the Higgs cubic and quartic self-interactions.
Because the Higgs boson has the largest couplings to the heaviest
Standard Model particles, promising Higgs production modes at colliders
are in association with $W$ or $Z$ bosons 
[$V^*\to V\hsm$ or $VV\to \hsm$] or 
$t$ quarks [$f\bar f,gg\to t\bar t\hsm$].
At hadron colliders, the largest production process
is $gg\to \hsm$, which is formally one-loop, but is enhanced
by the large gluon density in the proton.
At an $e^+e^-$ linear collider (LC), the dominant production mechanisms
are $e^+e^-\to Zh$ and $\to\nu_e\bar\nu_e h$.
Specialized colliders 
using muon beams ($\mu$C) or
photon beams ($\gamma$C) overcome
small couplings to $\hsm$ with beams tuned near the resonance
energy.
SM Higgs boson decays are dominated by the heaviest, kinematically
accessible particles.  Thus, for $\mhsm\lsim 135$ GeV,
the largest SM decay is $\hsm\to b\bar b$, but other sizable decays
in this mass range are $\hsm\to gg,\tau\tau,c\bar c,$ and $W^*W^*$.
Despite the dominance of only a few decay channels, it is nonetheless
important to calculate all decay rates to high precision.
For example, even though $\BR(\hsm\to\gamma\gamma)$ is typically 
${\cal O}(10^{-3})$,
it may be the easiest decay mode to observe at the LHC for a light
Higgs boson, 
because of a very clear
signal and excellent Higgs mass resolution. 
As $\mhsm$ increases, $\hsm\to WW^*$ then $\hsm\to WW+ZZ$ become the 
dominant decays,
even above the $\hsm\to t\bar t$ threshold.

\subsection{Discovery/Observation}

The best direct bounds on the SM Higgs boson mass
come from the CERN LEP collider, where searches for
the process $e^+e^-\to Z^*\to Z\hsm$ exclude 
$\mhsm<114.1$ GeV at the 95\% C.L.\cite{LEPHiggs:2001xw}. 
A slight ($2\sigma$)
excess of events 
has been observed near the kinematic limit,
which is consistent with $\mhsm=115.6$ GeV.
The indirect bounds from
precision Electroweak observables are consistent with this number,
and imply $\mhsm<196\gev$ at the 95\% C.L. \cite{LEPEW}.

\begin{figure}[!ht]
\resizebox{7cm}{6cm}{
\includegraphics[0,200][500,550]{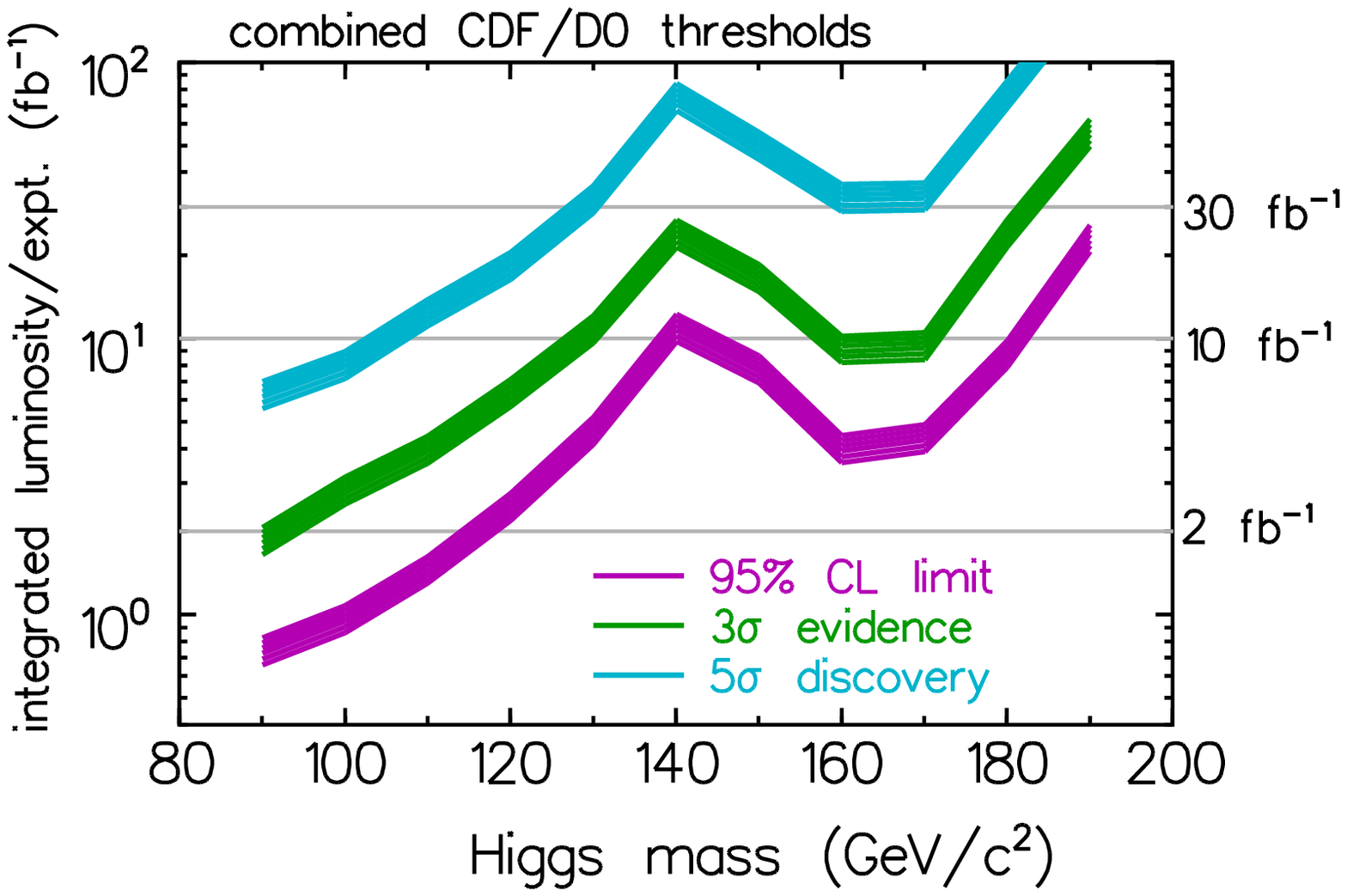}
}
\resizebox{7cm}{5cm}{
\includegraphics[0,0][560,460]{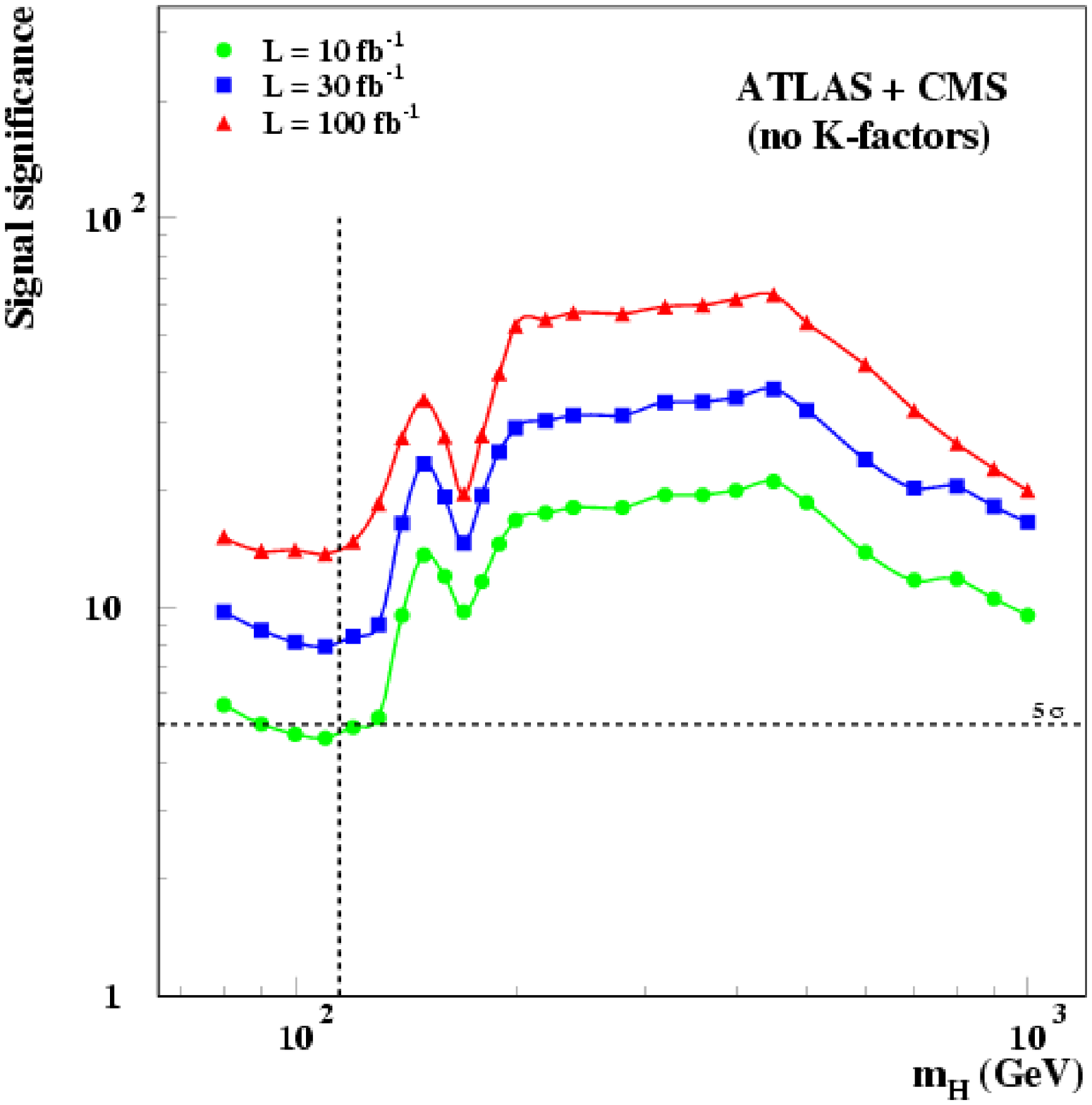}
}   
    \caption{(left) Required integrated luminosity to achieve different levels of
statistical significance for SM Higgs boson searches
at the Tevatron; (right) Statistical significance of SM Higgs boson
searches for different amounts of integrated luminosity at the LHC.}
    \label{fig:tevatron}
    \label{fig:lhc}
\end{figure}

The next collider experiments that are sensitive to the SM Higgs boson
are at the Tevatron $\rm p\bar p$ collider ($\sqrt{s} = 2$ TeV)
and the CERN Large Hadron $\rm pp$ Collider [LHC] ($\sqrt{s} = 14$ TeV).
At the Tevatron, the most accessible search channels are 
${\rm p\bar{p}} \to W\hsm+X$ and ${\rm p\bar{p}} \to Z\hsm+X$ with
leptonic gauge boson decays and $\hsm\to b\bar{b}$ \cite{TeVHiggs}. 
For $\mhsm>130$ GeV, 
the decay $\hsm\to WW^*$ can provide additional sensitivity.
Over the lifetime of the experiment, no single channel
can yield enough signal-like events to claim discovery.
However, the statistical combination of CDF and D\O~data
in all search channels has significant sensitivity.
The integrated luminosity needed for exclusion and 
discovery, respectively, are shown
in Fig.~\ref{fig:tevatron} as a function of the Higgs mass. With 
$2 (10)$ fb$^{-1}$ per experiment, $\mhsm < 120 (190)$ GeV can be excluded.
With 30 fb$^{-1}$ per experiment, discovery
at the 3--5$\sigma$ level can be achieved over the entire mass
range up to 180 GeV.
Roughly 2(5) fb$^{-1}$ will cover the region of the LEP excess
at 95\% C.L. ($3\sigma$).

At the LHC, there are at least three highly promising channels
for Higgs boson discovery: $gg \to \hsm \to \gamma\gamma$ for $\mhsm
\lsim 150 $ GeV and $gg \to \hsm \to ZZ^* \to 4\ell$ and 
$gg \to \hsm \to WW^* \to \ell^+\nu\ell^-\bar{\nu}$ for $\mhsm \gsim
130 $ GeV. Additional sensitivity comes from $gg,q\bar{q} \to
t\bar{t}\hsm (\hsm\to b\bar{b},\gamma\gamma)$ at low $\mhsm \lsim 120$ GeV 
and from 
$gg \to \hsm \to WW^*~{\rm or}~WW \to \ell\nu q\bar{q}$, 
both at the highest $\mhsm$ and also for $\mhsm\sim 160\gev$ where
the opening of the on-shell $WW$ channel suppresses the $ZZ^*$ signal.
With an integrated luminosity of 10 fb$^{-1}$ (out of a projected
$100$--$300$ fb$^{-1}$), discovery
at the $\approx$ 5$\sigma$ level is guaranteed over the whole
theoretically allowed mass range if information from these channels
and from both experiments,
ATLAS and CMS, are combined 
(see Fig.~\ref{fig:lhc}) \cite{Costanzo:2001ib,LHCHiggsI,unknown:1999fq}. 
Additional improvements are expected from the separate observation of 
weak boson fusion (WBF) channels, $qq\to qq\hsm$ with 
$\hsm\to\gamma\gamma$~\cite{Rainwater:1997dg},
$\hsm\to WW^*$~\cite{Rainwater:1999sd,Kauer:2000hi}, and
$\hsm\to \tau^+\tau^-$~\cite{Rainwater:1998kj,Plehn:1999xi}.

\begin{figure}[!ht]
  \begin{center}
\resizebox{7cm}{5cm}{
\includegraphics[136,456][476,691]{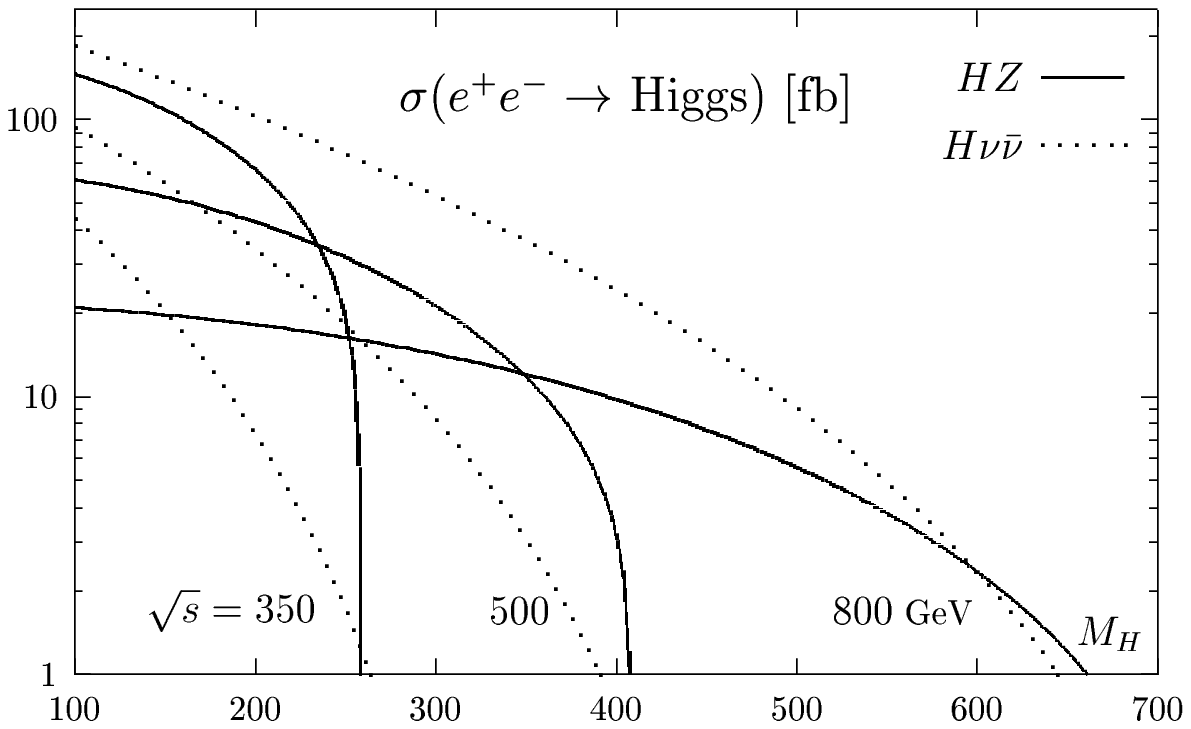}
}
\resizebox{7cm}{5.5cm}{
\includegraphics[0,0][566,566]{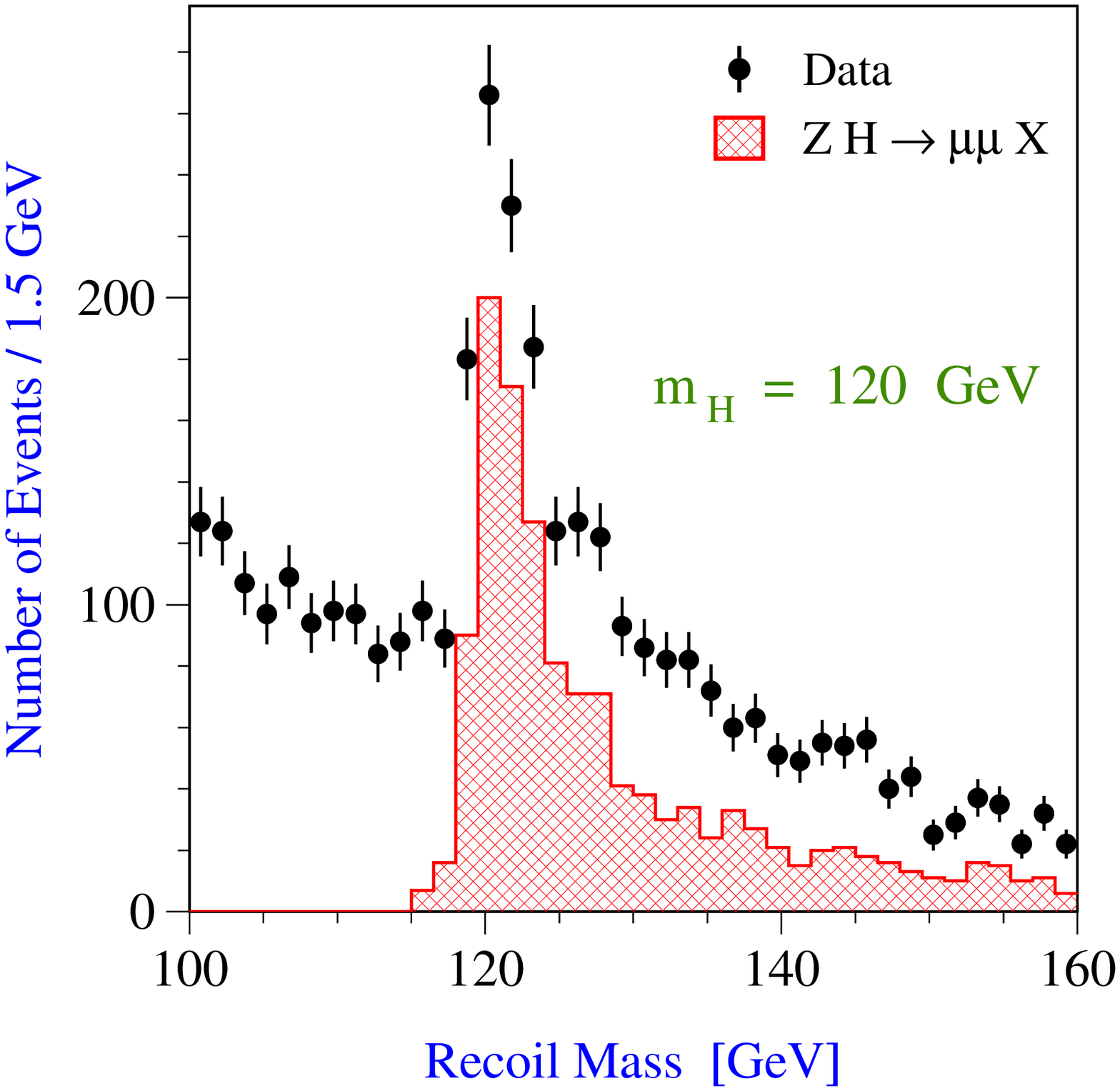}
}    
    \caption{ (left) Higgsstrahlung and $WW$-fusion cross sections for SM Higgs
boson production at a LC for various $\sqrt{s}$; (right) Recoil mass spectrum for $e^+e^-\to Zh$ with $Z\to\mu^+\mu^-$ signal and continuum background at a $\sqrt{s}=350$ GeV LC with 500 fb$^{-1}$ of
data and $m_h=120$ GeV.}
\label{fig:lcrecoil}
  \end{center}
\end{figure}

Proposed colliders, such as the 
LC, $\mu$C, and 
Very Large Hadron Collider (VLHC), would 
be SM Higgs boson factories.
The potential of a VLHC for performing precision Higgs studies is not
well--studied, and it is premature to make quantitative statements.
At the LC, the SM Higgs
boson is produced mainly in the $e^+e^- \to Z\hsm$ (Higgsstrahlung)
and $e^+e^-\to \nu_e\bar{\nu}_e \hsm$ (WBF) channels. 
Production rates for a light Higgs boson
are on the order of $10^5$ Higgs bosons for accumulated luminosity of
500 fb$^{-1}$ (corresponding to 1--2 years of running) 
at $\sqrt{s}\sim 350 $ GeV. The production cross
sections as a function of $\mhsm$ for various center--of--mass energies
are shown in Fig.~\ref{fig:lcrecoil}(a).
The Higgsstrahlung
channel offers the unique possibility to tag Higgs bosons independently
of their decay [Fig.~\ref{fig:lcrecoil} (b)], 
thus allowing for a model-independent observation 
of any Higgs boson with sufficient coupling to the $Z^0$.
In contrast to hadron colliders, Higgs bosons with any decay
mode can be selected at a LC with high efficiency $[{\cal O}(50\%)]$ 
and small backgrounds ($S/B\gsim 1$).

The LC is sensitive to a SM--like Higgs boson with a mass less than
or near the
kinematic limit.  If the Higgs coupling to $W$ and $Z$ bosons is
reduced from the SM value, the reach will be reduced accordingly.
The sensitivity to Higgs bosons with reduced couplings
to the $W$ and $Z$ bosons has not been well--studied, but the results
would be of great interest.

Observation of the Higgs boson in the recoil spectrum
could also be compromised if the Higgs is sufficiently wide to
wash out the signal.
This could occur if the Higgs has
a substantial decay width into pseudo--Goldstone bosons, 
additional Higgs singlets \cite{Binoth:1997au}, etc.
The exact sensitivity to such a case has also not been 
well studied.

The LC can be turned into a $\gamma$C by converting
beam electrons into highly energetic photons through back-scattering
of laser light. At a $\gamma$C, Higgs bosons
can be produced at resonance in the process $\gamma\gamma\to h$ with a large
cross section.  The one-loop $\gamma\gamma$ coupling is typically
large enough,
even for a Higgs boson with suppressed
tree-level couplings to the $W$ and $Z$ bosons,
that a $\gamma$C has
potential for discovering Higgs bosons that cannot be seen in $\epem$
collisions either by reason of couplings or mass.
The $\mu$C also allows the possibility of Higgs boson
production as an s-channel resonance. High rates are predicted
so long as the Higgs total width is not large. Especially useful
will be SM Higgs production for $\mhsm<180\gev$ and production of the 
$\hh,\ha$ of the MSSM \cite{Barger:1995hr,Barger:2001mi}.

\subsection{Properties:  Mass, Total Decay Width, Quantum Numbers,
and Couplings}
\subsubsection{Mass $\mhsm$}

The SM Higgs boson mass is fixed by the self-coupling $\lambda$,
which is constrained from 
above by perturbativity arguments [$\lambda^2/(4\pi)<1$],
and constrained from below by the requirement of vacuum stability 
[$\lambda>0$].
At the Tevatron, the statistical error on the Higgs mass measurement using
the $b\bar b$ invariant mass in $\hsm\to b\bar b$ decays will be
approximately $1$ GeV for $\mhsm=120$ GeV and $10$ fb$^{-1}$ of data.  
A conservative estimate including
systematic errors is 2 GeV, but the nearby $Z$ peak will be
quite useful as a calibration. 
At the LHC, for light Higgs bosons ($\mhsm \lsim 150$ GeV), the Higgs 
mass can be measured from the di--photon invariant mass in
$\hsm\to\gamma\gamma$ decays. The mass resolution is determined by
both the energy and angular resolution of the electro--magnetic
calorimeters in the ATLAS and CMS detectors. If an absolute 
normalization of the calorimeter energy scale of 0.1\% can be achieved, 
the mass resolution $\sigma_M/M$ 
is about 0.1--0.4 \% for 300 fb$^{-1}$. 
For larger Higgs masses, the $\hsm\to Z Z\to 4\ell$ decay provides 
similar precision.

At the LC, the Higgs mass is best reconstructed in the Higgsstrahlung
process either from the invariant mass recoiling against the $Z^0$ or
from a kinematic fit to the $Zh\to q\bar q b\bar b$ final state. 
The achievable precision is around 
5$\times 10^{-4}$ for $\mhsm = 120 $ GeV.  
For realistic operating scenarios, a $\gamma$C cannot provide any
further improvement of the accuracy of the Higgs mass determination.
Precision beyond the LC measurement can be obtained at the $\mu$C from a
scan of the $\mu^+\mu^-\to h$ resonance. Due to the expected excellent
control of the beam energy, a precision of roughly 10$^{-6}$ is 
envisaged.  

\subsubsection{Total Width $\gtot$}

The total decay width of the SM Higgs boson is predicted to be below
1 GeV for $\mhsm < 200 $ GeV, which is too small to be resolved 
directly except at the $\mu$C.
However, indirect methods can be employed both at the LHC and the LC.

\begin{figure}[thb]
\vspace*{1cm}
\begin{center}
\includegraphics[width=9.0cm,angle=90]{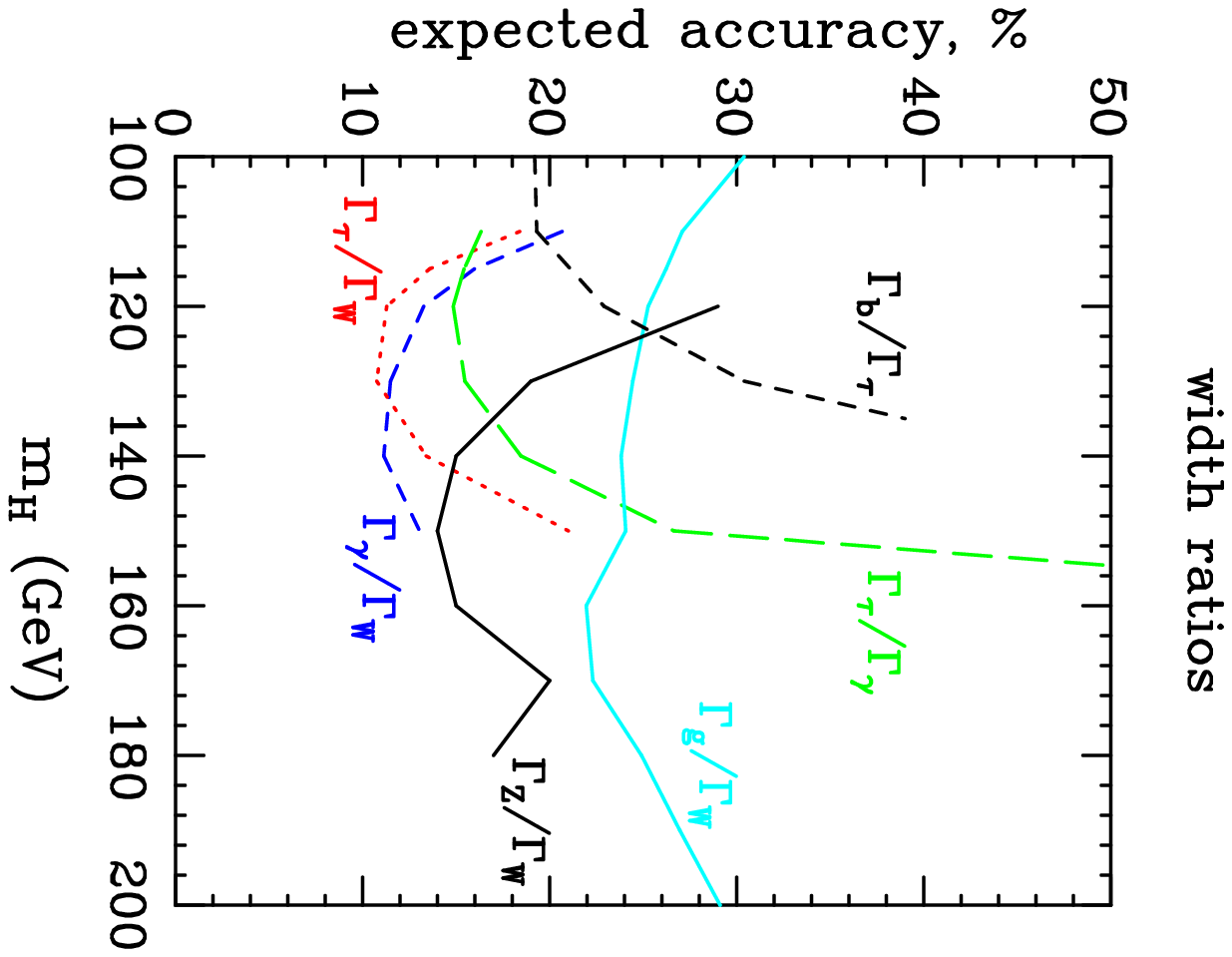}
\includegraphics[width=9.0cm,angle=90]{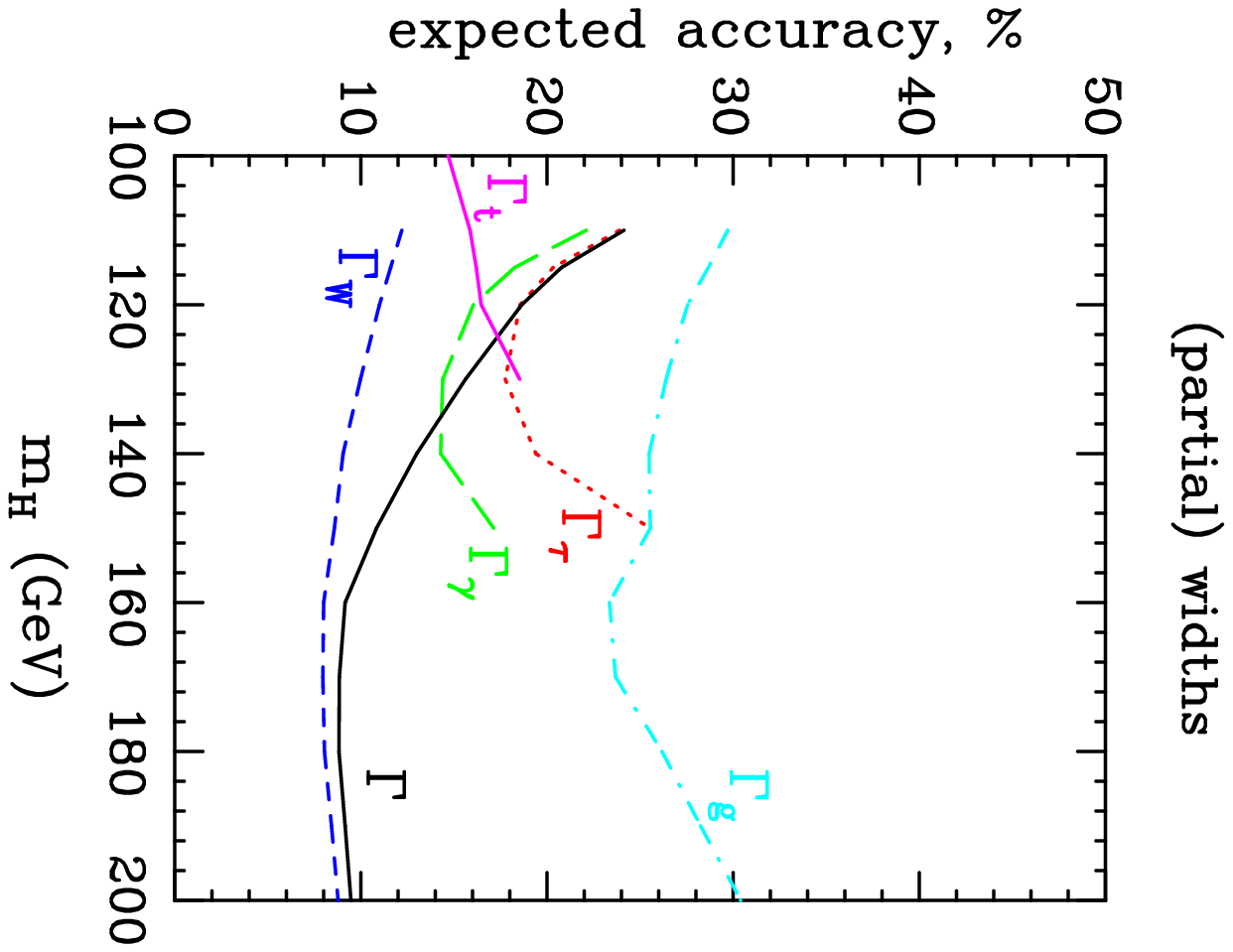}
\end{center}
\vspace*{0.2cm}
\caption{Relative accuracy expected at the LHC with 200~fb$^{-1}$ of data for
(a) various ratios of Higgs boson partial widths and (b) the indirect 
determination of partial and total widths $\tilde\Gamma$ and 
$\tilde\Gamma_i=\Gamma_i(1-\epsilon)$.  Simulations have been performed
at the parton level for WBF processes.
Width ratio extractions only assume $W,Z$ universality, which can be tested 
at the 15 to 30\% level (solid line). Indirect width
measurements assume $b,\tau$ universality in addition and require a
small branching ratio $\epsilon$ for unobserved modes like $H\to c\bar c$ 
and decays beyond the SM.}
\label{fig:Gamma}
%\vspace*{0.2cm}
\end{figure}

At the LHC, a variety of combinations of (partial) widths can be measured
directly. A few additional assumptions, which are appropriate for a 
SM-like Higgs, then allow the total width to be 
extracted~\cite{Zeppenfeld:2000td}.
From the weak boson fusion
processes $q\bar{q}\to q\bar{q} h$ with $h \to W W^*$, $
h\to\gamma\gamma$, and $h\to\tau\tau$, the quantities
$X_W = \Gamma^2_W / \Gamma$, $X_\gamma=\Gamma_W\Gamma_\gamma/\Gamma$, and
$X_\tau = \Gamma_W\Gamma_\tau / \Gamma$ can be measured.
The gluon--fusion induced
processes $gg \to h \to \gamma\gamma, ZZ^*, WW^*$ provide measurements
of $Y_i = \Gamma_g\Gamma_i / \Gamma$ for $i = \gamma,W,Z$. If
$SU(2)$ invariance holds for the $WWh$ and $ZZh$ couplings, if the ratio 
$y = \Gamma_b/\Gamma_\tau$ has its SM value and if the partial widths
of decay channels other than $ZZ, WW, b\bar{b}, \tau^+\tau^-, gg, 
\gamma\gamma$ are small, then $\Gamma_W$ can be
approximately determined from the quantities $X_W, X_\tau, X_\gamma, Y_Z, 
Y_W$ and $Y_\gamma$. Thus, the total width can be reconstructed 
under this assumption as 
$\Gamma = \Gamma_W^2 / X_W$. The achievable accuracy is shown 
in Fig.~\ref{fig:Gamma} and is estimated to
be about 20 (10) \% for $\mhl = 120 (200)$ GeV.
Despite the assumptions made, this method is useful, since the observation
of a width that substantially differs from the SM value
(predicted for a given $\mhsm$) would indicate new physics.
Another valid consistency check of the SM is to simply use
%the $f\bar f\to f'\bar f'h(\to WW)$ rate 
the rate for Higgs production in WBF, with subsequent decay $\hsm\to WW^*$, 
to infer the total width assuming
that $g^2_{WWh}$ and, hence, $\Gamma(h\to WW)$ have their SM values.
The latter is a good approximation for the lightest Higgs boson
of extensions of the SM Higgs sector when in the decoupling limit.

At the LC, the total  width of a SM-like $\hl$ can be computed in a 
model-independent manner.
For $m_{\hl}\gsim 120\gev$, the best method
is to first measure both
inclusive $\epem\to Z\hl$ production and $\epem\to Z\hl(\to WW^*)$
and compute BR($\hl\to WW^*$) 
from the ratio. The rate for $\epem\to\nu\anti\nu\hl(\to WW^*)$
determines $g_{WW\hl}^2\br(\hl\to WW^*)$ so that $g_{WW\hl}$ can be
extracted and $\Gamma(\hl\to WW)$ computed. 
The total $\hl$ width is then obtained in
a model-independent manner
as $\Gamma_{\rm tot} = \Gamma_W / BR(\hl\to WW)$.
For $\mhsm = 120$ GeV, an accuracy
of about 5\% can be achieved \cite{Desch:2001}.
Alternatively, the measurement of
BR($\hl\to \gamma\gamma$) at the LC can be combined with the measurement of
$\Gamma(\hl\to \gamma\gamma)$ at the $\gamma$C, yielding a somewhat
larger error on the total width $\sim 20\%$, 
due to the limited statistics of the
BR($\hl\to \gamma\gamma$) measurement.

Note that new physics contributions need not contribute
democratically to production and decay processes, so that
some ambiguity may exist in indirect extractions of the
Higgs boson width due to higher--order corrections.  
For example, within the MSSM, certain
box diagram contributions to the $e^+e^-\to Zh$ production process
that are absent for decay can modify the inclusive cross section
by 10\% \cite{Heinemeyer:2001iy} for certain choices of soft--breaking
parameters.  
Therefore, the effective $g_{ZZh}$ coupling would
not be the same one used in the calculation of the partial width.
It remains to be seen whether such corrections
can modify the $WW$--fusion process with a similar magnitude,
or whether the choice of soft--breaking parameters would 
necessarily provide a sparticle signature at the same or other
colliders.  While loop effects appear to be a complication,
an alternative view is that precise measurements of
the Higgs-strahlung cross section will provide additional information
about MSSM parameters \cite{Dawson:2002wc}.

For $\mhsm\gsim 200$ GeV, the total width can be
directly obtained from resolving the Higgs boson line-shape. At the 
LHC, the 4-lepton invariant 
mass spectrum from $\hsm\to ZZ\to 4\ell$ yields a precision of about 25\% for 
$\mhsm = 240 $ GeV, improving to about 5\% at 400 GeV and then slightly
degrading again~\cite{unknown:1999fq}.
At the LC, preliminary results at $\mhsm = 240 $ GeV 
show, that from a kinematic fit to the $\hsm Z\to ZZZ,WWZ$ final states with
one $Z$ decaying into a charged lepton pair and the other gauge bosons
decaying hadronically, a precision of about 10\% on the total width
can be obtained \cite{Meyer:2002}.

At the $\mu$C, the total width can be measured directly 
from a scan of the Higgs line-shape.
The expected accuracy for the $\hsm$ is of order $20\%$ at $\mhsm\sim 120\gev$
(i.e. poorer than the indirect LC technique but comparable to the LHC). 
For masses $\mhsm> 180\gev$, the $\mu$C does not yield an 
observable $\mu^+\mu^-\to \hsm$ signal~\cite{Barger:1995hr}.

\subsubsection{Quantum Numbers $J^{PC}$}

The SM Higgs boson has $J^{PC}=0^{++}$.
The observation of the $h\to\gamma\gamma$ decay
(e.g., at the LHC) would rule out $J=1$ and require C=$[+]$. 
At the LC, the spin
of the Higgs boson can be determined unambiguously
by examining the threshold dependence of the 
Higgsstrahlung cross section and the angular 
distributions of the $Z$ and Higgs bosons and their
decay products in the continuum \cite{Miller:2001bi,Garcia:2001wi,Hagiwara:1994sw,Barger:1994wt,Hagiwara:2000tk,Han:2000mi}.
Using optimal observables, a CP-odd component of a Higgs boson
coupling with strength $\eta$ relative to
an essentially SM-like CP-even component can be distinguished
at the $|\eta|\sim 3\%-4\%$ level.  Of course, in conventional
Higgs sector scenarios (such as the general 2HDM or
the MSSM with complex--valued soft--breaking parameters) $\eta$ is
roughly given by the fraction of the Higgs boson that is CP-odd
$f_{CP-}$ times a one loop factor, resulting in a value of $|\eta|$ that is too
small to be detected.  
However, in more general scenarios (in particular,
ones in which there are anomalous sources of $ZZ$ couplings 
such as those that can arise in composite Higgs models), $\eta$
could be of measurable size.  
The azimuthal distribution of the tagging jets in weak boson fusion
observed at the LHC is also sensitive to the appearance of non--renormalizable
CP--odd (and CP--even) operators \cite{Plehn:2001nj}.  
A simplified analysis implies sensitivity
at the $|\eta|\sim 30\%$ level.

At the LC, the (presumably dominant) CP--conserving production
processes $Z^*\to Zh, WW\to h$ can provide CP
information through correlations between the $\tau$ decay products
in the decays $h\to\tau^+\tau^-$ \cite{Grzadkowski:1995rx,Grzadkowski:1994kv,Kramer:1994jn}, but
this is challenging experimentally \cite{Bowertalk}.

In the $\gamma\gamma$ collision mode, the polarization of the photons can
be tuned to select different CP components of the
Higgs boson \cite{Grzadkowski:1992sa,Gunion:1994wy,Kramer:1994jn} and a reasonably good
determination of the CP nature of any Higgs boson observable
in this mode is possible. In the case of a SM--like
Higgs boson with $\mhl=120\gev$, a highly realistic NLC 
study \cite{Asner:2001ia}
concludes that a CP-odd component $f_{CP-}$ of about $20\%$
can be excluded at the 95$\%$ C.L. after one 
year of operation for expected luminosities.

At the $\mu$C, the transverse polarization of the muon beams can
be adjusted to select different CP components \cite{Barger:1997jm}. 
After accounting
for polarization precession there is some dilution of the transverse
polarization, but a good measurement will 
still be possible \cite{Grzadkowski:2000hm}.
For reasonable assumptions about proton source intensity and bunch merging,
one year of running will yield $b_\mu/a_\mu\lsim 0.2-0.3$ at the $1\sigma$
level assuming that in actuality $a_\mu=1$ and $b_\mu=0$, 
where $a_\mu$ and $b_\mu$
are the CP-even and CP-odd couplings of the Higgs boson to the muon
defined by $ {\cal L}_{\rm int}=-{gm_\mu\over 2 m_W}\anti \mu(a_\mu+ib_\mu\gamma_5)\mu\hl$.

Another promising method for directly observing CP violation is
to study the kinematics of Higgs bosons produced in association
with fermions as influenced by the relative sizes of $a_t$ and $b_t$
in the $t\anti t\hl$ coupling 
${\cal L}_{\rm int}= -{gm_t\over 2 m_W}\anti t(a_t+ib_t\gamma_5)t\hl$.
The reactions $p\bar p\to t\bar t h$ and $e^+e^-\to t\bar t h$
are sensitive to the CP nature of the Higgs $\hl$ at the 
LHC \cite{Gunion:1998hm} and LC \cite{Gunion:1996bk,Gunion:1996vv},
respectively.
Theoretical analyses find that for $\mhl\sim 120\gev$
the value of $b_t$ relative to the SM value of $a_t=1$ can be measured at about
the $40\%$ level at the LHC (using the $\gamma\gamma$
Higgs decay mode and for $300\fbi$ per detector)
and with an accuracy of $\sim 20\%$ at the LC 
with $\sqrt s=1\tev$ and $L=500\fbi$. Detailed experimental analyses
are not yet available.

Finally, for any type of Higgs sector, 
an incontrovertible 
signature of CP violation would be the observation of
2  separate, neutral Higgs bosons in $f\bar f\to Zh_1$, $f\bar f\to Zh_2$
{\it and} $f\bar f\to h_1h_2$ \cite{Gunion:1997aq}.  
Of course, this requires $\sqrt{s}>m_{h_1}+m_{h_2}$, while the
study of $t\bar th$ requires $\sqrt{s}>2m_t+m_h$.

\subsubsection{Gauge and Yukawa Couplings}

For a light Higgs boson, the Tevatron can observe two separate
production channels with the same Higgs boson decay.
The number of $Zh(\to b\bar b)$ and $Wh(\to b\bar b)$ 
final states could test the ratio $g^2_{ZZh}/g^2_{WWh}$ to
about 40\% for $\mhsm=120$ GeV and 10 fb$^{-1}$.
The LHC is sensitive to many
different production and decay processes, which provide
direct measurements of the ratios of several partial decay widths. The expected accuracy 
of such measurements is given in Fig.~\ref{fig:Gamma}, for 
100~fb$^{-1}$ of data in each of the two detectors~\cite{Zeppenfeld:2000td}. 
This corresponds to several years of running at lower luminosities,
where pile--up effects are not very important. 
For several channels, significant improvements can be expected  
with higher integrated luminosities.
However, a complete analysis of all channels, including pile--up at 
${\cal L}=10^{34}$cm$^{-2}$sec$^{-1}$, is not available at this time. 

The measurement of $\Gamma_b/\Gamma_\tau$ indicated in Fig.~\ref{fig:Gamma}
is assumed to originate from the 
$W\hsm(\to b\bar b)$~\cite{Drollinger:2002uj}
and $qq\to qq\hsm(\to \tau\tau)$ channels, where the former relies only on the 
CMS analysis for 300~fb$^{-1}$. A QCD uncertainty of 
10\% on the ratio of production cross sections is added in quadrature.
A better handle on 
$\hsm\to b\bar b$ decays is expected from $t\bar t\hsm(\to b\bar b)$ events
\cite{unknown:1999fq,Drollinger:2001ym} which provide a measurement 
of the combination 
$\Gamma_t\Gamma_b/\Gamma$ with a statistical accuracy of $12-14\%$ (for 
$m_{\hsm}<130$~GeV and 200~fb$^{-1}$). This is smaller than the 
NLO cross section uncertainty, which is taken to be 
10\%~\cite{Beenakker:2001rj}.
The dependence on the unknown top-Yukawa 
coupling $g^2_{htt}\propto \Gamma_t$ can be eliminated in principle,
by assuming top-quark
dominance in the $\hsm gg$ triangle graphs or by measuring 
$t\bar t\hsm$ production with subsequent decay $\hsm\to\gamma\gamma$ or 
$\hsm\to WW^*$~\cite{Maltoni:2002jr}. 

A global fit, using these techniques, is not available yet. Instead, 
tau--bottom universality is assumed for the extraction
of Higgs Yukawa couplings at the LHC~\cite{Zeppenfeld:2000td} and
a conservative 7\% error is assigned to the predicted $\BR(h\to\tau^+\tau^-)/\BR(h\to b\bar b)$
ratio.  
Expected 
accuracies for squared couplings, or, equivalently, (partial) decay widths, 
are given in Fig.~\ref{fig:Gamma}~\cite{dieter_snow}. 
For $\mhl=120$~GeV they are compared with LC expectations in 
Table~\ref{tab:g^2}. The errors on $\Gamma_t\propto g_{htt}^2$ and $\Gamma_g$
are dominated by systematics, namely QCD uncertainties at NLO and NNLO of
15 and 20\% for the $tt\hsm$ and $gg\to\hsm$ cross sections.

A LC can significantly improve these measurements
in a model--independent way.
The expected experimental uncertainties in the measurement of BRs at the 
LC for a 120 GeV SM-like Higgs boson are summarized in Table~\ref{tab:BRmeas}.
The first row shows the results assuming
500 fb$^{-1}$ of integrated luminosity at 
$\sqrt{s}=350$ GeV \cite{BattagliaDesch}.
The second row of Table~\ref{tab:BRmeas}
shows the results of a similar study \cite{Brau} for the branching
ratios of a 120 GeV SM-like Higgs boson with
500 fb$^{-1}$ at 
$\sqrt{s} = 500$ GeV.
Note the very different predictions 
in Table~\ref{tab:BRmeas} for the
precisions of $\BR(c)$ and $\BR(g)$, which depend on very good charm
and light quark separation.
The origin of this difference is not yet fully understood, and is not simply
a result of using different collider energies.
Finally, the entry in the last row is based
on the results of a dedicated study of the
$\BR(\gamma)$ measurement \cite{Boos}
for $\sqrt{s} = 350$ and 500 GeV, both without and with beam polarization
(80\% left-handed electron polarization and 40 or 60\% right-handed positron
polarization) chosen to enhance the Higgsstrahlung and $WW$ fusion cross
sections.
At $\sqrt{s} = 500$ GeV and the highest polarizations,
a measurement of $\BR(\gamma)$ with an experimental uncertainty of
9.6\% is possible with
1 ab$^{-1}$.  Scaling this to 500 fb$^{-1}$ to compare with the other
studies yields a precision of about 14\%, as shown 
in the third row of Table~\ref{tab:BRmeas}.
Without beam polarization, this deteriorates
to 16\% (23\%) with 1 ab$^{-1}$ (500 fb$^{-1}$).

\begin{table}[!ht]
\resizebox{8cm}{1.3cm}{
        \begin{tabular}{l c c c c c c}
        Decay mode: & $b \bar b$ & $W W^*$ & $\tau^+ \tau^-$ & $c \bar c$
                & $gg$ & $\gamma \gamma$ \\
        \hline
        Ref.~\cite{BattagliaDesch} & 2.4\% & 5.1\% & 5.0\% & 8.5\%
                & 5.5\% & 19\% \\
        Ref.~\cite{Brau}           & 2.9\% & 9.3\% & 7.9\% & 39\%
                & 18\% &  \\
        Ref.~\cite{Boos} (scaled)  &       &       &       &
                &      & 14\% \\
        \hline
        theory uncertainty  &  1.4\% & 2.3\% & 2.3\% & 23\%  & 5.7\%
                & 2.3\% \\
        \hline
        \end{tabular} 
}
\resizebox{8.8cm}{1.1cm}{
\begin{tabular}{l c c c c c c c c}
       &$g^2_{hWW}$  &  $g^2_{hZZ}$ & $g^2_{hbb}$ & $g^2_{h\tau\tau}$ & 
        $g^2_{hcc}$ & $g^2_{hgg}$ & $g^2_{htt}$ & $g^2_{h\gamma\gamma}$ \\
        \hline
LHC     &
11\%  & 11\%   & --    & 19\%  & --    & 28\%  & 16\% &  16\% \\
LC expt & 
2.4\% &  2.4\% & 4.4\% & 6.6\% & 7.4\% & 7.4\% & 10\% & -- \\
\hline
theory &
   -- & --     & 3.5\% & --    & 24\%  & 3.9\% & 2.5\% & -- \\
\hline
\end{tabular}
}
\caption{(left) Expected fractional uncertainty of BR measurements
at an $e^+e^-$ LC for a 120 GeV SM-like Higgs boson.
Results are shown for (500 fb$^{-1}$ at
$\sqrt{s} = 350$ GeV) (first row); 
(500 fb$^{-1}$ at $\sqrt{s} = 500$ GeV)
(second row); and (1 ab$^{-1}$ at $\sqrt{s} = 500$ GeV,
scaled to 500 fb$^{-1}$) (third row).  The theoretical uncertainty of
the predicted Standard Model branching ratios is given in the fourth
row.
(right) Expected uncertainty of measurements of squared couplings
(equivalently partial widths) for a 120 GeV SM-like Higgs boson. LHC results
correspond to Fig.~\ref{fig:Gamma}. LC estimates are from 
{\sc HFitter} \cite{BattagliaDesch,HFITTER}, 
assuming 500 fb$^{-1}$ at $\sqrt{s} = 500$ GeV, except for the
measurement of $g^2_{htt}$ which assumes  1 ab$^{-1}$ at
$\sqrt{s} = 800$ GeV.  The last line shows the theoretical uncertainty.}
\label{tab:BRmeas}
\label{tab:g^2}
\end{table}

At the LC, the extraction of the absolute couplings of the Higgs boson
is straightforward.  The coupling $g_{hZZ}$ is inferred directly
from the production cross section using the recoil method (modulo
the comment above -- this is true at the tree level).
Furthermore, most other couplings can be inferred from BR measurements
once $\gtot$ has been extracted using BR($h\to WW$) and
$\sigma(\nu_e\bar\nu_e h)$.
The results 
of a $\chi^2$ minimization 
using {\sc HFitter} \cite{BattagliaDesch,HFITTER} 
are summarized in Table~\ref{tab:g^2}.
The coupling $g_{ht\bar t}$ can be measured indirectly from the LC measurements
of $h \to gg$ and $h \to \gamma\gamma$ if one assumes that non--SM
loop 
contributions are small compared to experimental uncertainties.
A direct measurement of $g^2_{htt}$ can be obtained from the 
$e^+e^- \to t\bar t h$ cross section \cite{Gunion:1996vv,Aguilar-Saavedra:2001rg,JusteMerino,Baer:1999ge}.
Such a measurement requires running at higher $\sqrt{s} = 800\,$--$\,1000$~GeV
in order to avoid kinematic suppression of the cross section;
the result in Table~\ref{tab:g^2} assumes 1000 fb$^{-1}$ at 
$\sqrt{s} = 800$ GeV.

Sources of theoretical uncertainty include higher order loop
corrections to Higgs decay rates not yet computed and parametric
uncertainties due to the choice of input parameters.  The largest
sources of uncertainty arise from the choice of $\alpha_s$, $m_c$
and $m_b$ used in the numerical prediction:
$\alpha_s= 0.1185\pm 0.0020$,
$m_c(m_c)= 1.23\pm 0.09$~GeV \cite{ej} and $m_b(m_b)=4.17\pm
0.05$~GeV \cite{Hoang:2000fm}. 
The variation of these
input parameters 
leads to the the theoretical fractional uncertainties 
for the Higgs branching ratios quoted in 
Table~\ref{tab:BRmeas}.  For the Higgs squared-couplings listed in 
Table~\ref{tab:g^2}, the only significant theoretical uncertainties
reside in $g_{hbb}^2$ and $g_{hcc}^2$, due to the uncertainties in
the $b$ and $c$ quark masses and in $\alpha_s$ (which governs the
running of the quark masses from the quark mass to the Higgs mass).
The resulting theoretical uncertainties for $g_{hbb}^2$ and $g_{hcc}^2$ 
(for a SM Higgs boson of mass 120 GeV) are 3.5\% and 24\%, respectively.
In addition, a theoretical uncertainty in $g_{hgg}^2$ of 
3.9\% arises due to the uncertainty in $\alpha_s$.
The observed uncertainty in $m_t$ has only a small
effect on the predictions for the $h\to gg$ and $h \to \gamma\gamma$
decay rates.

For a SM Higgs boson with $m_h=120$~GeV, about 2/3 of the width is
due to $h\to b\bar b$.  The theoretical 
fractional uncertainties for the Higgs
branching ratios to $WW^*$, $\tau^+\tau^-$ and $\gamma\gamma$ 
listed in Table~\ref{tab:BRmeas} are
due primarily to the fractional uncertainty of the total width,
which for a SM Higgs boson with $m_h=120$~GeV is mainly governed
by the corresponding uncertainty in the $h\to b\bar b$ width.
For larger values of the Higgs mass, the $h\to b\bar b$
branching ratio is smaller and the uncertainty in the total width, 
which is now dominated by $h\to WW^{(*)}$, is
correspondingly reduced.
The large uncertainty in the $h\to c\bar c$ decay rate, arising
from the relatively large uncertainty in the charmed quark mass,
limits the usefulness of charm quark branching ratio and coupling
measurements.  
Further improvements in theory
and lattice computational techniques
\cite{charmlattice} may ameliorate the situation.

Finally, a scan of the $t\bar t$ threshold at the
LC will reduce the uncertainty on $m_t$ to
about 100~MeV \cite{Hoang:2000yr}.  Thus, the theoretical error
expected for the Standard Model Higgs coupling to $t\bar t$ due to the
top-quark mass uncertainty will be negligible.  The remaining uncertainty
in $g^2_{htt}$ is due to uncalculated higher order QCD corrections
to the $e^+e^- \to t \bar t h$ cross section.  We estimate this uncertainty
to be about 2.5\%
based on the renormalization scale dependence in the 
NLO QCD result for $m_h = 120$ GeV and $\sqrt{s} = 1$ TeV 
\cite{Dawson:1998ej,Dittmaier:2000tc}.

Another Higgs coupling that is potentially accessible in collider
experiments is $g_{h\mu\mu}$.  Of course, the $\mu$C relies on the
existence of such a coupling for $s$--channel production of the Higgs
boson.
The $\mu$C will allow a $\sim 2\%$ measurement of $g_{h\mu\mu}$ 
by computing $g_{\mu\mu \hl}^2\propto \sigma(\mu^+\mu^-\to\hl\to b\anti b)/\BR(b)$,
where $\br(b)$ is assumed to be measured with the above noted
precisions at the LC (see \cite{Barger:2001mi} and references therein).
The small theoretical error in the predicted
value of $g_{\mu\mu\hsm}$ would make this a particularly valuable
check of the SM expectation.
Recently, the prospects for measuring $g_{h\mu\mu}$ at other
colliders have been considered.  
A CLIC analysis claims an error for $g_{h\mu\mu}$ of $\sim 4\%$ from
WBF followed by $\hl\to\mu^+\mu^-$ \cite{Battaglia:2001vf}.  This latter
result relies
on an order--of--magnitude improvement in muon momentum resolution,
as assumed in many LC detector designs.  A precision of about $15\%$ would
be expected from operating at $\sqrt{s}=800$ GeV.
Based on a $5\sigma$ ``discovery'' in WBF with 300 fb$^{-1}$ of data,
a VLHC ($\sqrt{s}=200$ TeV, $\rm p\rm p$ collider) would provide a $10\%$ measurement \cite{Plehn:2001qg}.

\subsubsection{Self Coupling and Higgs Potential}

Further evidence of a fundamental scalar as the source
of
EWSB would be the reconstruction of the Higgs potential, i.e.
measurement of the self-coupling $\lambda$. 
The first kinematically accessible process with sensitivity
to $\lambda$ is $f\bar f\to Zhh$ \cite{Boudjema:1996cb,Ilin:1996iy}.
A {\bf full} reconstruction of the Higgs potential requires
measurement of $\mhsm$ (which is quite precise), the trilinear
coupling (which is good), and the quartic coupling.  Presently, there are no
claims for measurement of the quadratic coupling.  
The impact of this measurement can be phrased in the language of
non--renormalizable operators that will arise from a theory beyond
the SM which has a single, SM Higgs as its effective theory.
In this case, the Higgs potential can have the form:
\begin{equation}
V(\Phi) = \lambda (\Phi^\dagger\Phi-v^2)^2 + {\cal C}{(4\pi)^2\over \Lambda^2}
(\Phi^\dagger\Phi-v^2)^3 + \cdots.
\end{equation}
Expanding $\Phi$ into the physical Higgs component $v+H$, the
first term determines the coefficient of $H^2$ or the mass term, 
which we define
as $\lambda_{H^2}v^2$.  The first and second terms contribute to the
coefficient of the $H^3$ term, which we define as $\lambda_{H^3}v$.
In the renormalizable theory, $\lambda_{H^2}=\lambda_{H^3}$.
However, the LC measurement can only constrain this relation
at the 20\% level, i.e. $\lambda_{H^3}=(1.0\pm .2)\lambda_{H^2}$, which
bounds the relation
$\lambda_{H^3}=\lambda_{H^2}+8{\cal C}{(4\pi)^2v^2\over \Lambda^2}$.
Assuming a Higgs boson $m_H$ that is equal to $v$ 
and ${\cal C}=1$, the expected precision will
be sensitive to $\Lambda\lsim 20$ TeV.
In the MSSM, the measured self--coupling must obtain the
Standard Model value once the decoupling region is reached.
Some past studies have emphasized that the existence of
the trilinear coupling for the light Higgs boson will be
{\it verifiable} over most of the MSSM parameter space, but,
for the most part, it will not deviate from the Standard Model
expectations \cite{Djouadi:1999ei,Djouadi:1999gv}.
It is worth noting that some models of strong dynamics 
predict a light, composite
Higgs boson that also has a decoupling limit \cite{Dobrescu:1999gv}.
Therefore, the reconstruction of the Higgs potential would not
be incontrovertible evidence of a fundamental scalar.

\section{SUPERSYMMETRIC HIGGS BOSONS}

The Standard Model, with the Higgs mechanism
manifested by a single, fundamental
Higgs doublet that acquires a vacuum expectation
value, is believed to be an incomplete
description of nature.  There is only
a limited region of Higgs boson masses where
the theory remains perturbative through a 
desert up to a very high energy scale.  
Even in this limited region, the theory is extremely
fine-tuned.  
Furthermore, the apparent existence of a GUT structure,
as implied by the apparent unification of the
gauge bosons near the GUT scale, raises the question
of how a stable hierarchy can be maintained between
the Electroweak and GUT scales.  

One appealing solution
to the fine--tuning and hierarchy problems is the 
existence of a broken, Electroweak-scale
supersymmetry.  The simplest form of supersymmetry requires
an extended Higgs sector with two Higgs doublets, one
responsible for the masses of up-quark-like fermions [$H^0_2$]
and one for down-quark-like fermions [$H^0_1$].  The Higgs mechanism
then generates 5 physical Higgs bosons, labeled
$h, H$ $[CP=+]$, $A$ $[CP=-]$, and $H^+,H^-$.  Furthermore, the
Higgs boson self-coupling is no longer a free parameter,
but is proportional to (the square of) gauge couplings.
This implies a calculable upper bound to one of the Higgs
bosons of the theory.   

\subsection{Is a light $h$ supersymmetric in origin?}

To answer this question, we must understand the expected properties
of Higgs bosons in theories of Electroweak-scale supersymmetry (Susy)
breaking.
The properties of the Higgs sector are influenced primarily by
several soft-Susy-breaking mass parameters with values of ${\cal O}$(1 TeV)
and the Susy-conserving, dimensionless parameter
$\tanb$.  Lacking a clear picture of the Susy-breaking mechanism, we
vary these parameters over their allowed ranges, subject to
theoretical and experimental constraints, and examine the
consequences.  

The properties of the SM-like Higgs bosons can be derived from
the squared-mass matrix of the CP-even neutral MSSM Higgs bosons $h$
and $H$ [$m_{h} < m_{H}$], which is given in the $(H^0_1,H^0_2)$ basis by:
\begin{equation}
        \mathcal{M}^2 \equiv
        \left( \begin{array}{cc}
        \mathcal{M}^2_{11} & \mathcal{M}^2_{12} \\
        \mathcal{M}^2_{12} & \mathcal{M}^2_{22}
        \end{array} \right)
=
\left( \begin{array}{cc}
        m_A^2 s^2_{\beta} + m_Z^2 c^2_{\beta}
                & -(m_A^2 + m_Z^2) s_{\beta} c_{\beta} \\
        -(m_A^2 + m_Z^2) s_{\beta} c_{\beta}
                & m_A^2 c^2_{\beta} + m_Z^2 s^2_{\beta}
        \end{array} \right) +\delta{\cal M}^2.
        \label{eq:massmatrix}
\end{equation}
The contribution
$\delta{\cal M}^2$ is a consequence of the radiative corrections that
depend on the SM and Susy parameters.
Solving the eigen-problem for this matrix yields the physical masses
$m_h$ and $m_H$ and the mixing angle $\alpha$, which appears in the
Higgs couplings to fermions and gauge bosons.
The tree level prediction $m_{h} \leq m_Z |\cos 2\beta | \leq
m_Z$ is essentially ruled out by searches at LEP2.
However, the radiative corrections raise the
theoretical upper bound on $m_{h}$ substantially above $m_Z$.  
For a fixed value of $\tan\beta$
and a specified set of MSSM parameters,
$m_{h}$ grows with increasing $m_A$ and
reaches an asymptotic value 
$m_{h}^{\rm max}(\tan\beta)$.
If $\tan\beta$ is now
allowed to vary while holding all other free parameters fixed, 
$m_{h}^{\rm max}(\tan\beta)$ increases with
$\tan\beta$ and typically 
reaches an asymptotic value for
$\tan\beta \gsim 10$.  
For large values of $\tan \beta$, 
$m_{h} \simeq m_{h}^{\rm max}$ and
$m_{H} \simeq m_A$ for $m_A >
m_{h}^{\rm max}$.   Conversely, if $m_A < m_{h}^{\rm max}$ then
$m_{h} \simeq m_A$ and $m_{H} \simeq m_{h}^{\rm max}$.  
Based on a scan of MSSM parameters, the largest value
obtained for any reasonable choice of MSSM parameters
is $m_h^{\rm max}\lsim 135$ GeV.
Observation of a SM--like Higgs boson heavier than this would
rule out a simple [or even CP--violating] MSSM Higgs sector.  In the
NMSSM, which includes an additional Higgs singlet, $m_h^{\rm max}$ can
be increased from its MSSM value for some choice of parameters.
The inclusion of
the dominant two--loop contributions to the effective potential have
reduced the previous upper--bound to around 
$135-140$ GeV \cite{Yeghian:1999kr,Ellwanger:1999ji}.
However, based on using a similar approximation in the MSSM, one
can expect a further shift of several GeV after including sub--leading
contributions.

Must such a Higgs boson be observed in the next generation of collider
experiments?  This will be the case if the light MSSM Higgs boson 
has substantial tree level coupling to $W$ and $Z$ bosons.
In the MSSM, this follows from the fact that
the couplings of $h$ [$H$] to the $W$ and $Z$ are given by
$\sin(\beta - \alpha)$ [$\cos(\beta - \alpha)$] times the corresponding
SM Higgs coupling and from the CP-even Higgs boson
sum rule \cite{Carena:2000bh,EspCom,Espinosa:1998xj}
\begin{equation}
        m_{H}^2 \cos^2(\beta - \alpha)
        + m_{h}^2 \sin^2(\beta - \alpha)
        = \left[m_{h}^{\rm max}(\tan\beta) \right]^2\,.
\end{equation}
In particular, if $m_A<m_h^{\rm max}$ and $\tanb$ is large
one finds $m_H\sim m_h^{\rm max}$
and $\cos(\beta-\alpha)\sim 1$ while for $m_H\gg m_h^{\rm max}$
(as occurs in the $m_A\gg m_Z$ decoupling limit) the sum rule
requires $m_h\sim m_h^{\rm max}$ and $\sin(\beta-\alpha)\sim 1$
(for any $\tanb$).

Will this Higgs boson be sufficiently different to exclude the
SM?
To answer this question, we must address the decoupling limit in
more detail.
In the decoupling limit, we find that 
$\sin(\beta-\alpha)=1$ [or equivalently  
$\cos(\beta-\alpha)=0$], in which case the couplings of $\hl$ are
identical to those of the $\hsm$.  
This behavior, which is easy to verify for the
tree-level expressions, continues to hold when radiative corrections
are included.  However, the onset of decoupling can be significantly
affected by the radiative corrections.  In general,
\bea \label{eq:cosbma}
\cos(\beta-\alpha) = {(\mathcal{M}_{11}^2-\mathcal{M}_{22}^2)\sin 2\beta
-2\mathcal{M}_{12}^2\cos 2\beta\over
2(m_H^2-m_h^2)\sin(\beta-\alpha)} 
= {{m_Z^2\sin 4\beta+
({\delta\mathcal{M}}_{11}^2-{\delta\mathcal{M}}_{22}^2)\sin 2\beta
-2{\delta\mathcal{M}}_{12}^2\cos 2\beta}\over
2(m_H^2-m_h^2)\sin(\beta-\alpha)}\,.
\eea
Since $\delta\mathcal{M}^2_{ij}\sim {\mathcal O}(m_Z^2)$, 
and $m_H^2-m_h^2=m_A^2+\mathcal{O}(m_Z^2)$, one obtains 
\begin{equation} \label{cosbmadecoupling}
        \cos(\beta-\alpha)=c\left[{m_Z^2\sin 4\beta\over
        2m_A^2}+\mathcal{O}\left(m_Z^4\over m_A^4\right)\right]\,;\,\,
\label{cdef}
        c\equiv 1+{{\delta\mathcal{M}}_{11}^2-{\delta\mathcal{M}}_{22}^2\over
        2m_Z^2\cos 2\beta}-{{\delta\mathcal{M}}_{12}^2\over m_Z^2\sin
        2\beta}\,;
\end{equation}
Eq.~(\ref{cosbmadecoupling}) exhibits the expected decoupling behavior 
for $m_A\gg m_Z$, but also reveals 
another way in which
$\cos(\beta-\alpha)=0$ can be achieved---Nature must simply choose the
supersymmetric parameters (that govern the Higgs mass radiative
corrections) such that $c$ vanishes.  Remarkably, the vanishing of
$c$ is independent of $m_A$, and has a large $\tan\beta$ solution at
\begin{equation} \label{earlydecoupling}
\tan \beta\simeq \frac{2m_Z^2-
\delta\mathcal{M}_{11}^2+\delta \mathcal{M}_{22}^2}
{ \delta\mathcal{M}_{12}^2}\,.
\end{equation}
Explicit solutions depend on ratios of Susy parameters and  
so are mostly insensitive to the overall Susy mass scale.

The behavior
of the MSSM Higgs couplings as the decoupling limit
is approached is revealed by expressing them in terms of $c$:
\begin{equation}
   {g^2_{hVV}\over g^2_{h_{\rm SM}VV}}=\sin^2(\beta-\alpha)\simeq 1-{c^2 m_Z^4\sin^2 4\beta\over
4m_A^4}\,,        \label{eq:g2wwdecoupling}
\end{equation}
which quickly assumes the SM value as $m_A$ increases.
At large $\tan\beta$, the approach to decoupling is even faster, since
$\sin 4\beta\simeq -4\cot\beta$.

The couplings of $h$ to up-type fermions may 
be written (explicitly for $t$):
\begin{equation}
{g^2_{htt}\over  g^{2}_{h_{\rm SM}tt}  } = 
\left[ \sin(\beta - \alpha) + \cot\beta \cos(\beta - \alpha)
        \right]^2
\left[ 1 - \frac{\Delta_t\tan\beta}{1 + \Delta_t} 
        (\cot\beta + \tan\alpha) \right]^2 
        \simeq 1+\frac{c m_Z^2\sin 4\beta\cot\beta}{m_A^2}
\label{eq:g2ccdecoupling}
\end{equation}
where the couplings are expressed in terms of $\sin(\beta - \alpha)$
and $\cos(\beta - \alpha)$ in order to better illustrate the decoupling
behavior.  The Susy vertex corrections, expressed as $\Delta_t$, 
are absent from the final expression since they
are not enhanced with $\tan\beta$ and the prefactor
$\cot\beta+\tan\alpha \simeq \cos(\beta-\alpha)/\sin^2\beta$ is small in
the decoupling limit.
Similar expressions can be written for the charm quark, in which
case the Susy vertex corrections are entirely negligible.
The approach to decoupling is significantly
slower [by a factor of $m_A^2/m_Z^2$] than in the case of the $hVV$ coupling
[Eq.~\ref{eq:g2wwdecoupling}].  
At large $\tan\beta$, the approach to decoupling is faster due to the
additional suppression 
factor of $\cot^2\beta$ as in the case of the $hVV$ coupling.

For the coupling of $h$ to down-type fermions,
Susy vertex corrections cannot be neglected.
Focusing on the $b\bar b$ coupling,
and neglecting corrections
which are not $\tan\beta$-enhanced, it follows that
\begin{equation} 
  {g^2_{hbb}\over g^2_{h_{\rm SM}bb}}
        = \left[ \sin(\beta - \alpha) - \tan\beta \cos(\beta - \alpha)
        \right]^2 \left[ 1 - {\Delta_b\cot\beta\over 1+\Delta_b}(\tan\beta+\cot\alpha)
\right]^2
\simeq 1-{4c m_Z^2\cos 2\beta
\over m_A^2}\left[\sin^2\beta-{\Delta_b\over 1+\Delta_b}\right]\,.
        \label{eq:g2bbdecoupling}
\end{equation}
The approach to decoupling is again slower as compared to $g_{hVV}$.
However, in contrast to the previous two cases, there is no
suppression at large $\tan\beta$.  In fact, since
$\Delta_b\propto\tan\beta$, the approach to decoupling is further
delayed, unless $c\simeq 0$.  A similar expression can be written for
the $\tau$ coupling.  The function $\Delta_\tau$ is also $\tan\beta$ enhanced,
but is of order $g^2$ instead of $g_s^2$ and $y_t^2$, and is thus expected to
be of less importance.

\subsection{Coverage of Susy parameter space with the Light Higgs Boson}

Sensitivity to the supersymmetric origin of a light Higgs boson
from Higgs coupling measurements depends on the closeness to
the decoupling limit.
Since Higgs widths are approximately quadratic in the
couplings, the greatest
deviations from the SM are expected in $\Gamma(b)$ and
$\Gamma(\tau)$, since these quantities approach the decoupling
limit slowly.
Determining the ``coverage'' of Susy parameter space is a biased
endeavor.  It is commonly phrased in terms of coverage in
the $m_A-\tan\beta$ plane for fixed values of all other soft-
Susy-breaking parameters.

The projected reach of the LHC for a Standard Model Higgs boson
implies that a light, CP-even Higgs boson as predicted by the MSSM
will also be observable over the entire 
$m_A-\tan\beta$ plane. For $m_A\gsim 110\gev$, it is
the $\hl$ that is SM-like and which will be detected. For lower $m_A$,
where $\sin^2(\beta-\alpha)$ is suppressed,
the observed SM-like Higgs would be the heavier $H$.
The ability to always detect the SM-like Higgs boson survives even
when supersymmetric particles
are included in the one-loop $hgg$ and $h\gamma\gamma$ couplings.
This is because it is essentially impossible to simultaneously
suppress these couplings, and hence the $gg\to h\to \gamma\gamma$ rate,
while also suppressing the
$t\bar th(\to b\bar b)$ or $WW\to h(\to \tau^+\tau^-)$ processes
that, as has been discussed, 
provide very good signals in the SM Higgs case \cite{Carena:2000bh}.
While observation of a light, SM-like Higgs boson at the LHC 
will be consistent
with a supersymmetric origin, this alone would not
be incontrovertible evidence for Susy.  The existence of
sparticle--like signatures would be supporting, but not conclusive,
evidence.  

The LC has increased sensitivity to the properties of $h$ from BR
and coupling measurements.  
Nonetheless, even the LC will not 
be able to distinguish the light CP-even Higgs of the MSSM
from a Higgs boson with precisely SM-like properties
if the MSSM lies in the decoupling limit.  Furthermore,
this decoupling can occur much more rapidly than expected based purely
on dimensional analysis, i.e. $m_A \gg M_Z$.
If the MSSM parameters are such that the $m_A$-independent 
decoupling is realized, then the 
experimental sensitivity to $m_A$ is greatly compromised.
One way to present coverage of Susy parameter space is 
to consider ``benchmark'' scenarios
that lead to very
different behaviors of the SM-like Higgs boson of the MSSM.
Three such scenarios are summarized in Table~\ref{tab:scenarios}, and 
correspond approximately to those discussed in Ref.~\cite{Carena:1999xa}.
All MSSM parameters are specified at the electroweak scale.
The three benchmark scenarios have the following properties:

\begin{description}
\item[No-mixing scenario:]  The top squark mixing angle $\theta_{\tilde t}$
is zero.
This scenario yields the lowest
value of $m_{h}^{\rm max}(\tan\beta)$
for given values of $\tan\beta$ and $M_S$.
For simplicity, the scenarios are defined in terms of
$M_{\rm Susy} \equiv M_{\tilde Q} = M_{\tilde U} = M_{\tilde D}$,
where the latter are third generation squark mass parameters.
For $\msusy\gg m_t$, as is true in the scenarios considered here,
$M_{\rm Susy}\simeq M_S$ [where
$M^2_S \equiv .5(M^2_{\tilde t_1} + M^2_{\tilde t_2})$].
Here, a large value for $\msusy = 1.5$ TeV is chosen
in order to obtain
a sufficiently large value of $m_{h}^{\rm max}(\tan\beta)$,
comparable to that obtained in the other
two scenarios (the case of $\msusy=1$ TeV is at the edge of the region
excluded by LEP2).
\item[Maximal-mixing scenario:]  The top squark
mixing is chosen to
give the maximal value of $m_{h}^{\rm max}(\tan\beta)$
for given values of $\tan\beta$ and $M_S$.
\begin{table}[!ht]
    \begin{center}\begin{tabular}{c c c c c c l} \hline
        & \multicolumn{5}{c}{Mass parameters [TeV]} & \\
        Benchmark & $\mu$ & $X_t \equiv A_t - \mu \cot\beta$ & $A_b$
                & $M_{\rm Susy}$ & $M_{\tilde g}$ & $m_h^{\rm max}$ [GeV]\\
        \hline
        No-Mixing & $-0.2$ & 0 & $A_t$ & 1.5 & 1 & 118 \\
        Maximal-Mixing & $-0.2$ & $\sqrt{6}$ & $A_t$ & 1 & 1 & 129\\
        Large $\mu$ and $A_t$ & $\pm 1.2 $ & $\mp 1.2(1 + \cot\beta)$
                & 0 & 1 & 0.5 & 119 \\
        \hline
        \end{tabular} \end{center}
        \caption{MSSM parameters for our benchmark scenarios,
and the derived maximal mass for the SM-like Higgs boson.}
        \label{tab:scenarios}
\end{table}
\item[Large $\mu$ and $A_t$ scenario:]
Large radiative corrections occur to both the mixing angle $\alpha$ and
through $\Delta_b$.  In particular,
$\mathcal{M}_{12}^2$ can exhibit extreme variations in magnitude
depending on the sign of $A_t\mu$ and the magnitude of $A_t$.
The two possible sign combinations for $A_t$ and $\mu$ 
(for a fixed sign of $A_t\mu$)
yield small differences in
$\mathcal{M}_{12}^2$ through the dependence of $h_t$ and $h_b$
on $\Delta_t$ and $\Delta_b$, respectively.
The vertex correction $\Delta_b$ is dominated by the
bottom squark-gluino contribution, which can enhance or suppress
the Yukawa coupling $h_b$ for negative or positive $\mu$, respectively.
In the following, two possible sign
combinations for $A_t$ and $\mu$ are considered with $A_t\mu< 0$.
\end{description}

To be conservative, relatively large values for the Susy 
breaking parameters, on the order of 1 TeV, are chosen so that some supersymmetric
particles may not be kinematically accessible at the LC.
However, for simultaneously large $\mu$ and $M_{\tilde g}$, the size of the 
$\Delta_b$ corrections may drive the bottom Yukawa coupling out of the 
perturbative region.  
Thus the gluino mass is taken as $M_{\tilde g} = 0.5$ TeV for large $\mu$ and
$M_{\tilde g} = 1$ TeV for moderate $\mu$.  The other gaugino 
mass parameters are $M_2=2 M_1=200$ GeV 
($M_2$ is relevant for the one-loop $h \to \gamma \gamma$ amplitude).
Finally, the masses of the remaining squarks and sleptons 
are set to 1 TeV.

The LC studies summarized earlier for branching ratio
and coupling measurements
were conducted
for $\mhsm=120$ GeV, and thus are directly
applicable to the
study of a SM-like Higgs boson of the MSSM with a mass
near 120 GeV, especially near the
decoupling limit.  
Deviations from SM behavior can be probed using $\delta \BR\equiv|1-\BR_{\rm MSSM}/\BR_{\rm SM}|$ or $\delta \Gamma$ (defined similarly).  Since the
overall sensitivity is similar, only results for $\delta \BR$ are shown.

\begin{figure}
\resizebox{9.5cm}{!}{
\includegraphics*[0,142][472,682]{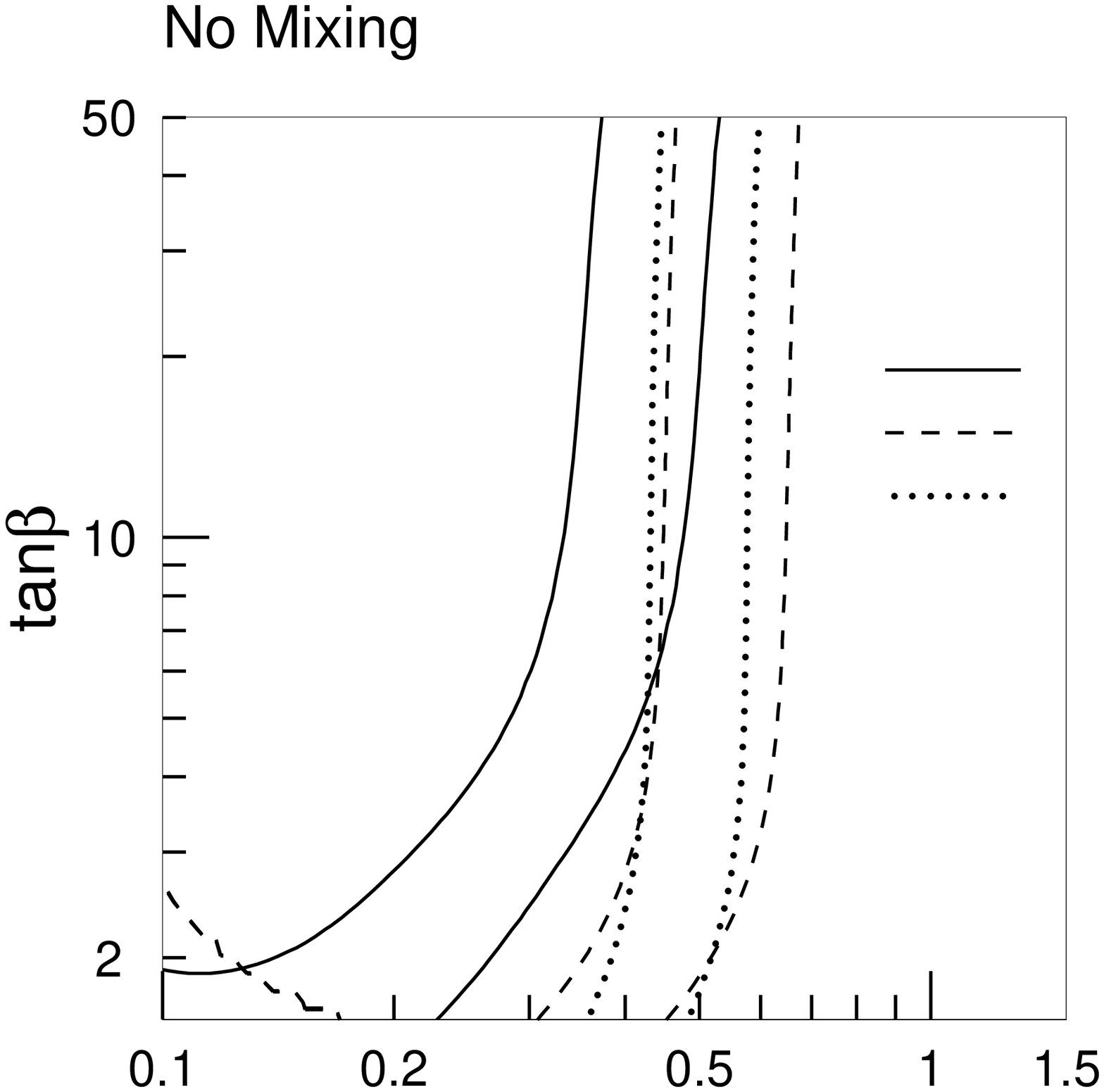}
\includegraphics*[57,142][529,682]{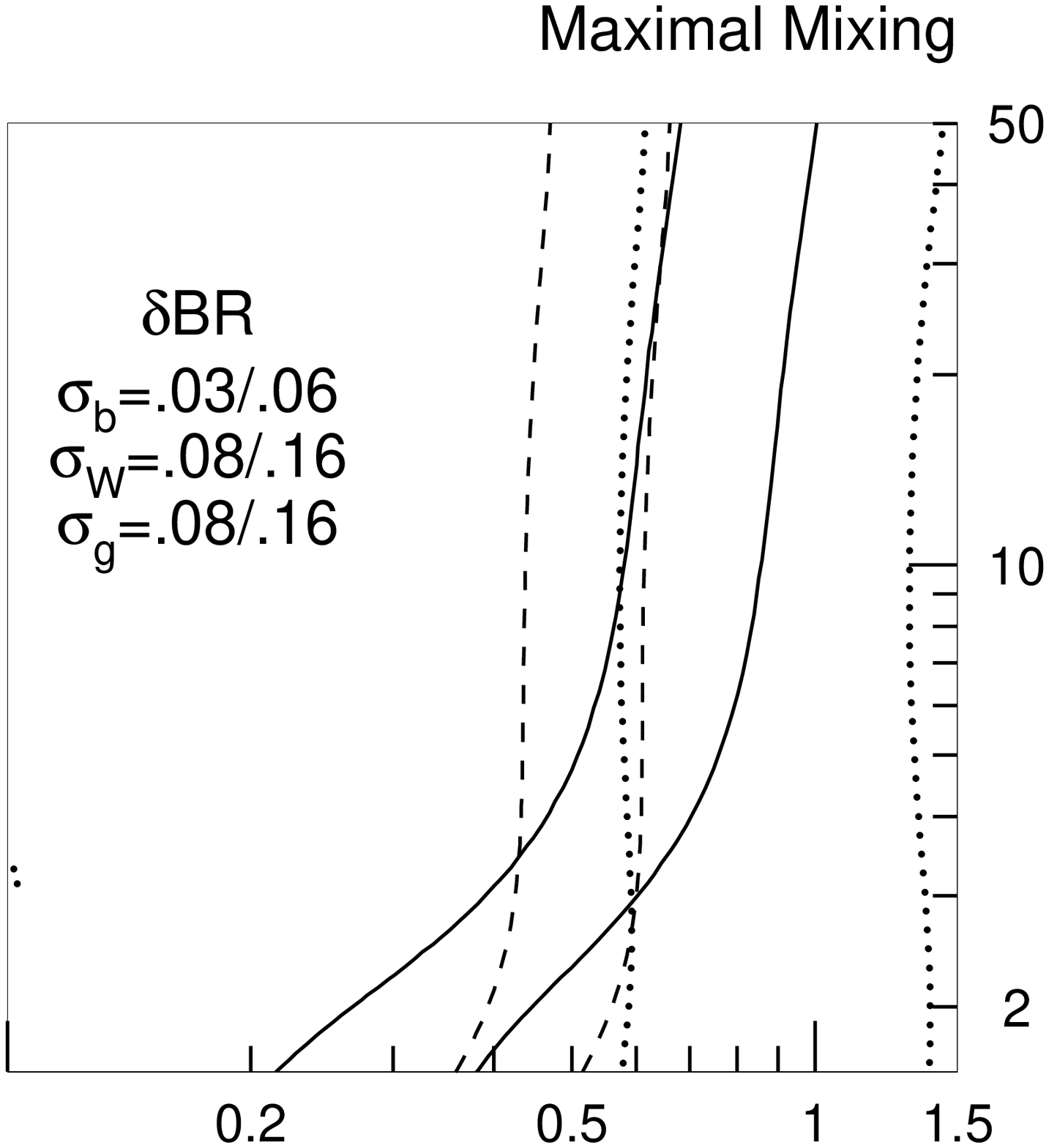}
}

\resizebox{9.5cm}{!}{
\includegraphics*[0,142][472,682]{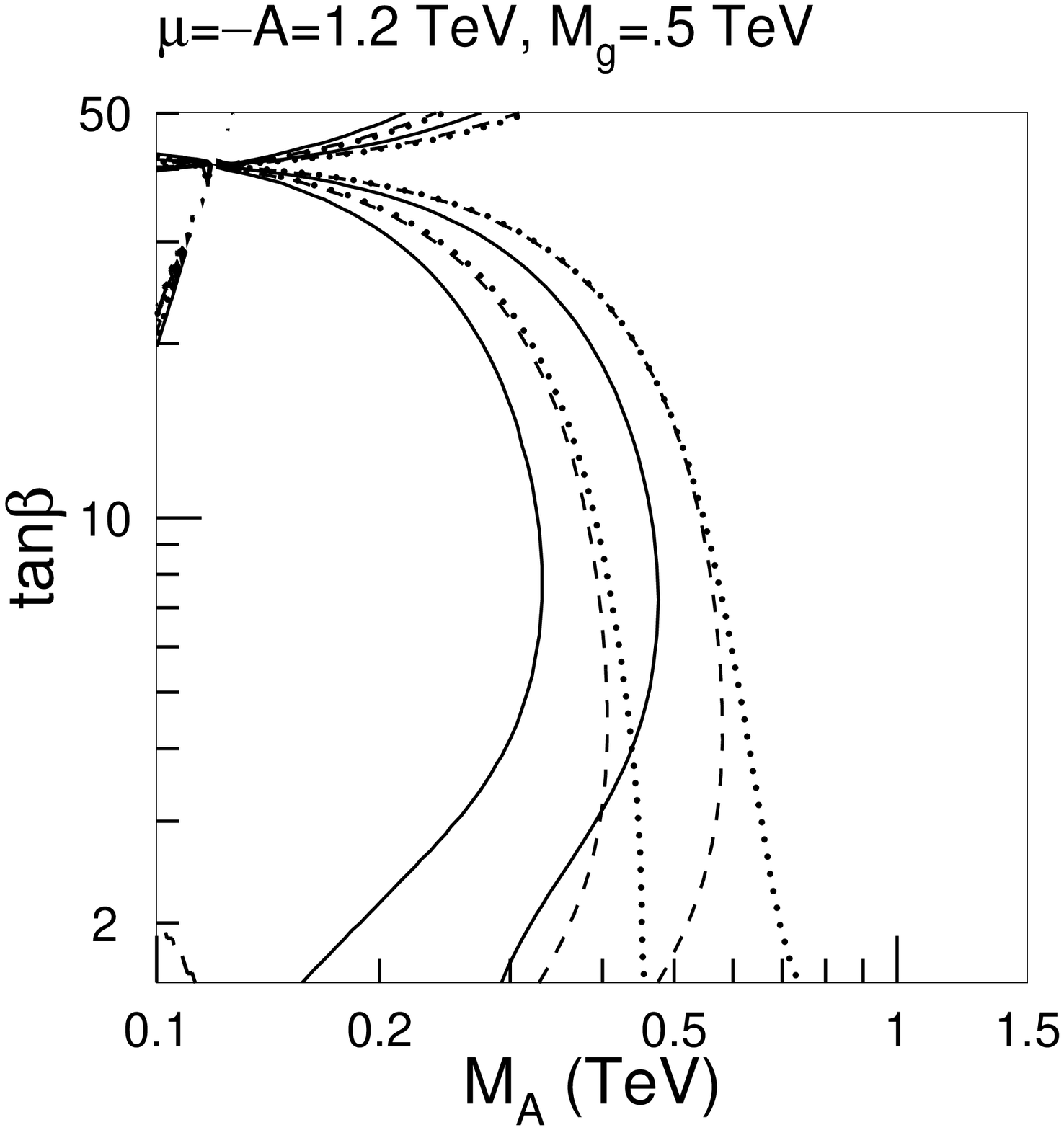}
\includegraphics*[57,142][529,682]{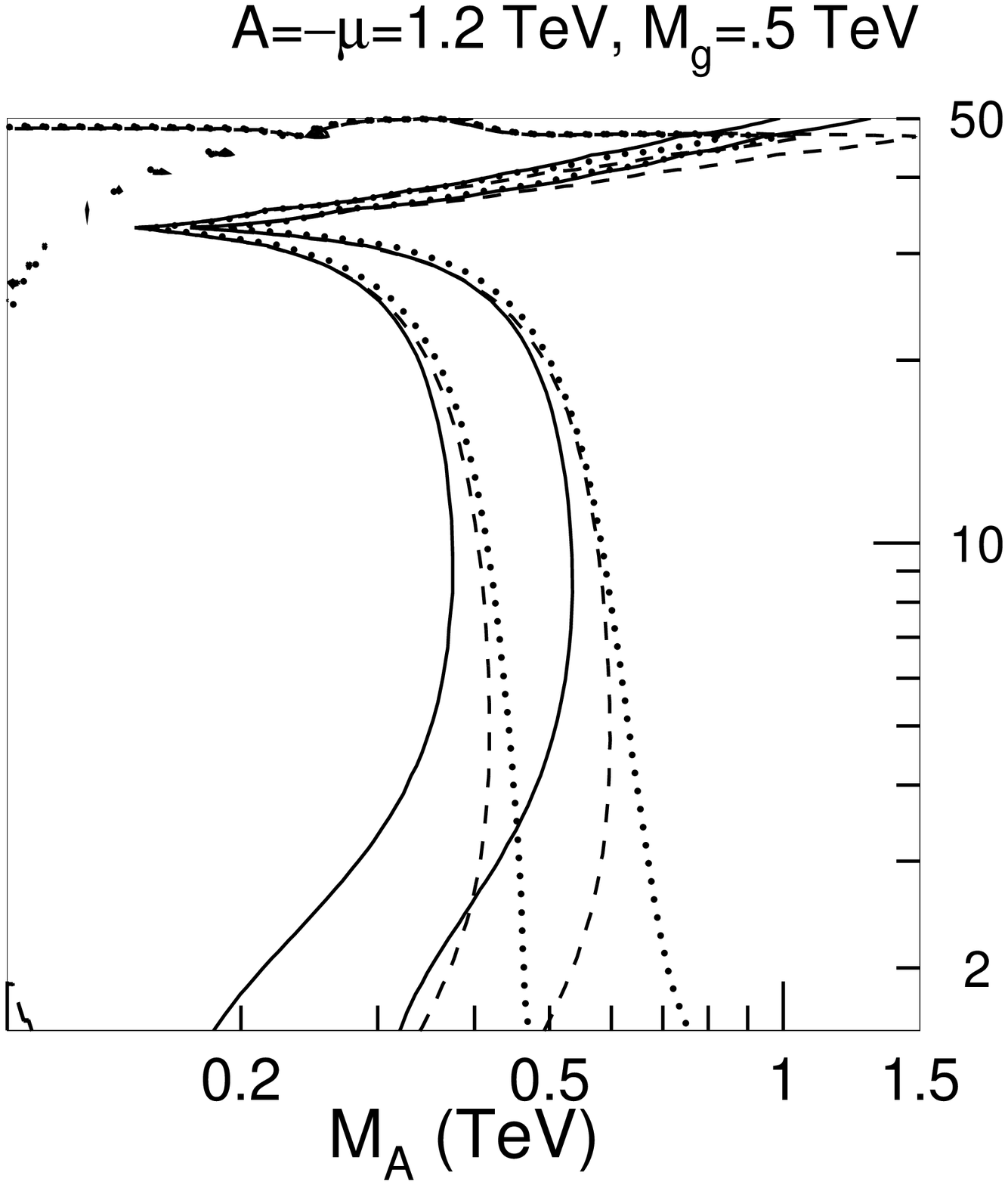}
}

\caption{Contours of
$\delta \BR(b) = 3$ and 6\% (solid),
$\delta \BR(W) = 8$ and 16\%
(long-dashed) and
$\delta \BR(g) = 8$ and 16\% (short-dashed)
in the three benchmark scenarios.}
\label{figure1}
\end{figure}

In the four panels of Fig.~\ref{figure1}, the solid, long-dashed, 
and short-dashed lines are
contours of $\delta\BR(b)$, $\delta\BR(W)$ and $\delta\BR(g)$, 
respectively \cite{Carena:2001bg}. 
Although $\delta \Gamma(b)$ is quite large
over much of the parameter space, $\delta \BR(b)$ is smaller because the
increase in $\Gamma(b)$ also significantly increases $\Gamma_{\rm tot}$.
Because $\delta \Gamma(W)$ quickly approaches zero for increasing $m_A$,
$\delta\BR(W)$ indicates variation
in the total Higgs width, and is more sensitive than $\delta\BR(b)$,
except for the case of maximal mixing.
In regions of parameter space 
where $\delta \Gamma(g)$ approaches zero, 
$\delta\BR(g)$, like $\delta\BR(W)$,
is sensitive to variations in the total width.

For the maximal-mixing scenario,
the mass of the SM-like Higgs boson near the decoupling limit is
roughly 10 GeV heavier than in the other benchmarks
(see Table~\ref{tab:scenarios}), so that
the relative contribution of $\Gamma(b)$ to $\Gamma_{\rm tot}$ is
decreased.  Therefore, deviations in $g_{hbb}$ are not as
diluted in the BR measurement as in the other scenarios,
and  the measurement of $\delta\BR(b)$ yields superior
sensitivity at large $\tan\beta$, around $m_A \lsim 600$--700 GeV at
$2\sigma$.  
One should interpret
this result with caution, however, since the accuracies for
BR measurements are based on the simulation of a 120 GeV
SM Higgs boson.
In the maximal-mixing scenario,
$\BR(g)$ deviates by more than 8\% from its SM value for
$m_A \lsim 1.4$ TeV.
At $2 \sigma$
the reach in $\delta \BR(g)$ is roughly $m_A \lsim 600$ GeV.
In the no-mixing scenario,
$\delta\BR(g)$ and $\delta\BR(W)$
give comparable
reach in $m_A$; at $2 \sigma$ the reach is $m_A \lsim 425$ GeV.
For comparison, in the no-mixing scenario 
deviations in $\BR(b)$ yield sensitivity at
$2 \sigma$ for $m_A \lsim 300$ GeV for $\tan\beta \gsim 5$.

In the large $\mu$ and $A_t$ scenarios,
$\BR(g)$ gives the
greatest reach in $m_A$, allowing one to distinguish the MSSM from the
SM Higgs boson at $2 \sigma$ for
$m_A \lsim 350$--450 GeV, depending
on the value of $\tan\beta$.
At larger values of $\tan\beta$, the large $\mu$ and $A_t$ scenarios
have regions of $m_A$-independent decoupling where the SM-like MSSM Higgs
boson cannot be distinguished from the SM Higgs boson even for very low
values of $m_A$.

Clearly, the regions of the $m_A$---$\tan\beta$ plane
in which the MSSM and SM Higgs bosons can be distinguished from one another
at the LC by measuring the properties of the light Higgs boson
depend strongly on the supersymmetric parameters, and the sensitivity
comes from different measurements for different sets of MSSM parameters.
In the following section, the impact of other Susy Higgs measurements
will be considered.

\subsection{Coverage of Susy parameter space and Heavy $H/A$ and $H^\pm$ Bosons}

In or near the decoupling limit,
measurements of the properties of the light Higgs boson in the MSSM
will most likely
not deviate significantly from the SM expectations.  
Of course, it is possible
that $H$ ($m_H < 135$ GeV)
is the SM-like Higgs boson and $hA$ production
occurs ($m_h < m_H$ and $m_h\sim m_A$), but this is far from the decoupling
limit.
When $h$ is very SM-like, it becomes 
urgent to observe one or more of the heavier Higgs bosons.
The relevant couplings of the neutral Higgs bosons are:
\begin{eqnarray*}
        g_{HVV}/g_{h_{SM}VV} = \cos(\beta-\alpha);\,\,\,g_{AVV}/g_{h_{SM}VV} = 0  
\end{eqnarray*}
\vskip -0.7cm
\begin{eqnarray*}
        g_{Htt}/g_{h_{SM}tt} = \cos(\beta-\alpha)+\sin(\beta-\alpha)\cot\beta;\,\,\,
        g_{Att}/g_{h_{SM}tt} = \cot\beta\, 
\end{eqnarray*}
\vskip -0.7cm
\begin{eqnarray}
        g_{Hbb}/g_{h_{SM}bb} = 
\left[ \cos(\beta-\alpha)+\sin(\beta-\alpha)\tan\beta \right]
        \frac{1}{1 + \Delta_b} \left[ 1 - \Delta_b \cot^2\beta \right];\,\,\,
        g_{Abb}/g_{h_{SM}bb} = 
         \tan\beta
        \frac{1}{1 + \Delta_b}\,. 
        \label{eq:couplingsDeltab}
\end{eqnarray}
There are similar expressions with $t\to c$ and $b\to \tau$.
As $\cos(\beta-\alpha)\to 0$, the $H$ boson couplings approach those
of the $A$ boson, which has no significant coupling to $W$ and $Z$
bosons.  While the true pseudoscalar still has $\gamma_5$
in its interaction Lagrangian, differences in the predictions
of production and decay properties for the $H$ and $A$ bosons
depend on the kinematic mass of the associated fermions, and are difficult
to observe.  Also, the effects of Susy vertex corrections can have
important effects on the $H$ and $A$ properties.
Finally, there is a $ZHA$ coupling proportional to $\sin(\beta-\alpha)$,
which is maximal in the decoupling limit.  

For the charged Higgs boson $H^\pm$, where there is no SM analog,
it is more convenient to
write the interaction Lagrangian between fermions and the charged Higgs boson:
\begin{eqnarray}
{\cal{L}} & = & 
   (h_b + \bar{\delta} h_b ) H^- \bar{b}_R t_L \sin\beta +
\bar{\Delta} h_b \cos\beta H^- \bar{b}_R t_L
\nonumber \\
& + & (h_t + \bar{\delta} h_t) H^+ \bar{t}_R b_L \cos\beta +
\bar{\Delta} h_t H^+ \bar{t}_R b_L \sin\beta + h.c.,
\label{Lag}
\end{eqnarray}
with similar expressions for $\tau-\nu_\tau$ and $c-s$.
The presence of the 
couplings 
$\bar{\delta} h_{b,t}$ and $\bar{\Delta} h_{b,t}$
\cite{Carena:2000yi} indicate deviations from the tree-level
expectations, and their effect is similar in nature to
the $\Delta_i=\Delta h_i/h_i$ corrections introduced earlier.
Finally, there are also $\gamma H^+H^-$ and $ZH^+H^-$ couplings,
as well as other couplings which are not important near the decoupling
limit.

\begin{figure}[htb]
\resizebox{9.5cm}{!}{
\includegraphics*{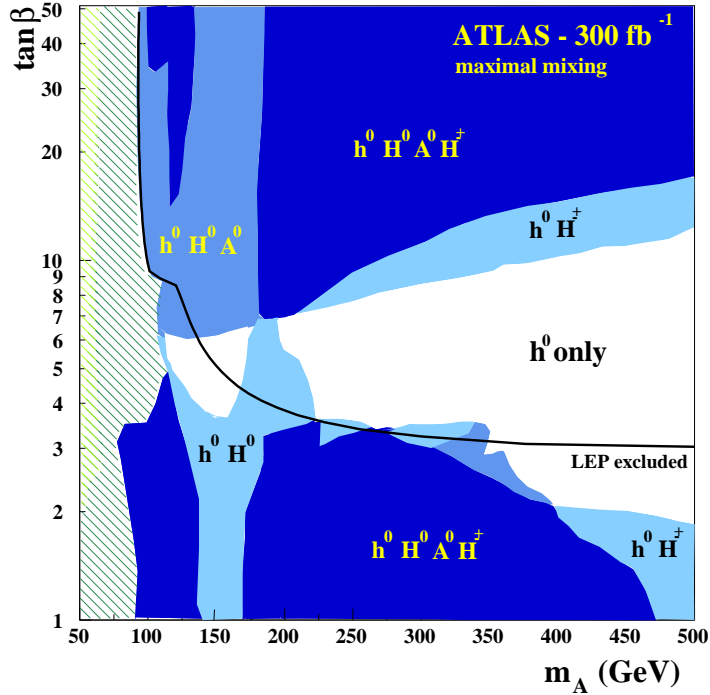}
}  
  \caption{LHC coverage of the $m_A-\tan\beta$ plane for a conservative Susy model, but neglecting potentially large corrections at large $\tan\beta$.}
  \label{fig:tdr2}
\end{figure}

The pseudoscalar $A$ boson can be produced at a substantial rate at
hadron colliders
either in association with $b$ quarks [$q\bar q,gg\to b\bar bA$
or $gb(\bar b)\to b(\bar b)A$] or through
$gg\to A$ when $b$ or $\widetilde b$ loops dominate, provided
that the parameter $\tan\beta$ is large enough.
Because of the large heavy-quark backgrounds, search strategies
have focused on the decays $\tau^+\tau^-$.  In much of Susy
parameter space, except for the region near $m_A\simeq m_h^{\rm max}$,
one of the CP-even Higgs bosons has very similar properties to
$A$ [except for CP], thereby increasing the expected signal 
(the present studies have not treated carefully the mass splitting
between $A$ and either $h$ or $H$ and how the displaced mass
peaks affect the estimate of the ``continuum'' background shape).
In the decoupling limit, the charged Higgs also has a mass similar
to that of $A$ and $H$, and can be produced in the process 
$gb\to tH^-$, for example.  
The expected coverage in the $m_A-\tan\beta$ plane at the LHC 
is displayed in Fig.~\ref{fig:tdr2}, for a conservative set of
Susy parameters and neglecting any Susy vertex effects, i.e.
$\Delta_b$ and $\Delta_\tau$.    
While a light SM--like Higgs boson is observable over the entire
unexcluded region, its properties can be indistinguishable from
a true SM Higgs boson, and thus shed no light on the underlying physics.
However, for $\tan\beta\gsim 10$, 
the $H^\pm, A$ and $H$ bosons can be 
discovered through production in association with $t$ or $b$ quarks and decays 
into $\tau$ leptons, with coverage deteriorating somewhat at larger values
of $m_A$.

The effect of $\Delta_b$ on the $b\bar b$
couplings of the heavy MSSM Higgs bosons does {\it not} decouple 
for $m_A\gg m_Z$ ({\it i.e.}, for 
$\tan\alpha\tan\beta = -1$).  Thus $\Delta_b$ could potentially
have a significant 
effect on the discovery of the heavy Higgs bosons at the Tevatron and 
LHC by modifying production cross sections and decay branching ratios.  
The full impact of the Susy vertex corrections for hadron colliders
has not been adequately explored; they will at least cause ambiguity
in the interpretation of a purported value of $\tan\beta$.
Here, we estimate their effect.
The hadron collider production rate for
$b\bar b \phi(\to b\bar b)+X$ 
scales roughly as $\tan^2\beta/(1+\Delta_b)^2$ as long as the $b$ decay
is still dominant, while the rate for
$b\bar b \phi(\to\tau^+\tau^-)+X$ is mainly unaffected and
scales as $\tan^2\beta$,
due to a cancellation between changes in the production rate
and changes to the total width.
Ignoring uncertainties from higher-order QCD
corrections and other sources, discovery of
the $H$ and $A$ will
allow an extraction of $\tan\beta/(1+\Delta_\tau)$ with a 
relative error of about 10\%.  
The impact on the charged Higgs discovery and interpretation should
be similar, since modifications to the production rate are similarly
canceled by the change to the total width.  
Studies described in the ATLAS TDR \cite{atlas_tdr} show
a relative error of 6\% (7\%) on the extraction of $\tan\beta$
from $H/A\to\tau^+\tau^-$ decays
assuming large $\tan\beta$, 300 fb$^{-1}$ of
data, and $m_A=150 (350)$ GeV.  The relative error decreases to
about $5-6$\% using the rarer decays to muon pairs.  
However, these methods are sensitive to any Higgs decays into
sparticles.

At a LC, the relevant
production processes for heavy Higgs bosons
are $(\gamma/Z)^*\to b\bar b (t\bar t) H/A$, $(\gamma/Z)^*\to H^+H^-$,
and $Z^*\to HA$.  The $b\bar b H/A$ process requires rather large
values of $\tan\beta$, whereas $t\bar t H/A$ requires small $\tanb$
and $\sqrt{s}>$350 GeV $+m_{H/A}$ \cite{Grzadkowski:1999wj,Djouadi:1992tk}.  
Small $\tan\beta$ is theoretically
disfavored, since it typically leads to a lighter SM-like Higgs boson
that should have been observed at LEP.
$HA$ $[H^+H^-]$ production is fairly independent of
$\tanb$, but requires $\sqrt{s}>m_H+m_A [2m_{H^\pm}]$.
Thus, for high enough $\sqrt{s}$, the LC can observe heavy Higgs bosons
in the moderate $\tan\beta$ region where the LHC does not have
$5\sigma$ sensitivity.  Note that the $5\sigma$ criterion is
quite stringent, but analyses of the 95\% CL and $3\sigma$ limits
possible at the LHC are not yet available.
The kinematic reconstruction of Higgs decay
products in pair production at the LC will also allow a fairly
good determination of the Higgs masses 
\cite{Kiiskinen:2000wv,Battaglia:2001be}, on the order of 1\%.

The large $\mu$ and $A_t$ scenario introduced above demonstrates the 
complementarity of the LC and the hadron colliders.
For $\tan\beta \lsim 10-20$, where the heavy MSSM Higgs bosons
can be missed at the LHC, $\BR(g)$ gives the
greatest reach in $m_A$, allowing one to distinguish the MSSM from the
SM Higgs boson at $2 \sigma$ for
$m_A \lsim 350$--450 GeV, depending
on the value of $\tan\beta$.
At larger values of $\tan\beta$, the large $\mu$ and $A_t$ scenarios
have regions of $m_A$-independent decoupling where the SM-like MSSM Higgs
boson cannot be distinguished from the SM Higgs boson even for very low
values of $m_A$.  
In fact, in these scenarios it is possible for $h$ to be indistinguishable
from the SM Higgs boson at the LC, while at the same time $m_A < 250$ GeV
so that the heavy Higgs bosons will be directly observed at a 500 GeV
LC through $e^+e^- \to HA$, $H^+H^-$. 
Where 
the $m_A$-independent decoupling occurs,
for $\tan\beta \simeq 40$ [$\tan\beta \simeq 33$],
the heavy MSSM Higgs bosons 
would be discovered at the LHC even for $m_A$ above 500 GeV \cite{atlas_tdr,unknown:1999fq}.
Note also that for large $\mu$ and $A_t$ and large $\tan\beta$,
the correction $\Delta_b$ is quite large.

A very accurate determination of $\tanb$, mainly
through Higgs production rates and Higgs decays,
is also possible at a LC \cite{Gunion:2001qy}.  This is important,
since it is fairly difficult to obtain a robust measurement of
this parameter from other observables at hadron colliders
or even a LC.
As an example, assuming ${\cal L}=2$~ab$^{-1}$, 
$\mha\sim\mhh\sim 200\gev$ and $\rts=500\gev$, 
measurements of the total $\hh$ and $\ha$ widths from
calorimetry alone will
yield accuracy $\Delta\tanb/\tanb\sim 0.1$ ($\sim 0.02$) for
$\tanb\sim20$ ($\sim 50$), while 
ratios of branching ratios yield $\Delta\tanb/\tanb<0.1$
for $\tanb\leq 13$; $\bb\ha+\bb\hh$ production yields 
$\Delta\tanb/\tanb<0.1$ for $\tanb\geq 40$.
These different techniques are highly complementary and will, in combination,
allow a very accurate determination of $\tanb$.  
Charged Higgs production should also be visible at the LC
when $HA$ production is, and this will provide additional information.
At the LC, the processes $e^+e^-\to\tau^+\nu H^-$ and $\to t\bar bH^-$
show some sensitivity, but the most promising channel is
$(\gamma/Z)^*\to H^+H^-$. 
%%% Already mentioned above.
%The charged Higgs boson mass can also 
%be measured to high precision. 
While these results are promising, 
the analysis also ignored the dependence on Susy vertex corrections.
The extracted value of ``$\tan\beta$'' only represents an effective
coupling.
At large $\tan\beta$, where only a few decay modes dominate, the vertex
corrections can have a significant impact.  From the above discussion,
however,
it appears that a combination of LHC and LC measurements can give
a determination of $\tan\beta$ {\bf and} the radiative corrections.
Clearly, LC data would be of great 
value for disentangling the $\Delta_b$ dependence, so that the value
of $\tan\beta$ extracted from heavy Higgs boson measurements can be compared
to the value obtained from other sectors of the theory.

At a $\gamma$C, the $(\gamma\gamma\to H)$+$(\gamma\gamma\to A)$ signal
is observable in the $b\anti b$ final state for many
$m_A-\tanb$ parameter choices.  In particular, for almost all of the 
moderate-$\tanb$ wedge region with $250\gev\lsim \mha\lsim 500\gev$
of Fig. \ref{fig:tdr2} in which only the light $h$ 
of the MSSM can be detected at the LHC (with a 
similar wedge of non-detection being present at a 
LC operating in the $\epem$ collision mode 
\cite{Grzadkowski:1999wj,Farris:2002ny}), 
the combined $(\gamma\gamma\to H)+(\gamma\gamma\to A)$
signal will be observable at the $4\sigma$ level \cite{Asner:2001ia} 
after about 3
years of operation at $\rts=630\gev$ using the NLC design. The
factor of two larger luminosity at TESLA would yield full
coverage of this wedge region after 3 years of operation.
(See also \cite{Muhlleitner:2001kw} and references therein.)
Of course, the LHC and LC  wedge regions in which
$H,A$ discovery will not be possible extend to arbitrarily
high masses beyond $500\gev$,
spanning an increasingly large range of $\tanb$ as $m_A$ increases.
Discovery of the $H/A$ signal at the $\gamma$C for masses above $500\gev$
would require higher energy for the electron beams at the LC;
roughly, $H/A$ masses $\lsim 0.8\rts$ could be explored --- a detailed
study is needed to determine the amount of luminosity required
to achieve 4 to 5 sigma signals throughout this entire kinematically
accessible mass range for all $\tanb$ in the
wedge region when operating at a fixed high $\rts$.

Finally, at a $\mu$C, direct observation of $H/A$ is possible through
the $\tan^2\beta$ enhancement in 
$\sigma(\mu^+\mu^-\to H/A)$ \cite{Barger:1997jm,Barger:2001mi}.
The strength of the $H,A$ signals at the $\gamma$C and $\mu$C
are sensitive to the $\Delta_b$ radiative corrections and could
be enhanced or suppressed relative to the tree-level expectations
employed in the above studies.

\subsection{CP violation}

In the MSSM, CP need not be a good quantum number once
one-loop effects are included in the Higgs potential.  Mixing
between the scalars and the pseudoscalar arises if there is
a phase mismatch between several combinations of soft Susy-breaking
parameters, i.e. if $arg(A_t\mu)\ne 0$ \cite{Pilaftsis:1999qt,Carena:2000yi}.
As a result, each of the three neutral Higgs bosons can have
a CP-even admixture, and, thus, couple to the $W$ and $Z$ bosons.
Observation of 3 separate Higgs bosons with couplings
to $W$ and $Z$ might be possible, depending upon the precise mixing,
the Higgs masses and (at the LC) the available energy.  
Also, all 3 of the couplings $Z\to h_1h_2$, $h_1h_3$ and $h_2h_3$
would be significant in general, allowing observation of
all these pair processes at a LC with sufficient 
energy.
As noted earlier, observation of $Z^*\to h_i Z$, $Z^*\to h_j Z$
and $Z^*\to h_ih_j$ for any $i\neq j$ is a direct signal of CP violation
(\cite{Grzadkowski:1999ye} and references therein).
CP-violating effects in the MSSM are most important
for $m_{H^\pm}\lsim 170$ GeV and 
$\tan\beta\lsim 7$ \cite{Carena:2000ks},
implying that all the Higgs boson mass eigenstates would typically 
have masses in this same range, making all 6 of the above processes
kinematically accessible at a $\rts\gsim 350-500\gev$ machine.   
Since $e^+e^-\to H^+H^-$ would be visible for $\rts=350\gev$
for $m_{H^\pm}\sim 150\gev$, we would be alerted to the possibility
of this scenario even at an early stage LC. 
Susy QCD effects can render such a charged Higgs invisible in top
quark decays or even in direct production at a hadron collider 
\cite{Carena:1999py}.

As sketched earlier, direct observation of CP violation 
via final state distributions in Higgs-strahlung production of 
a single Higgs boson is very
difficult. In particular, in the MSSM the 
coupling of the CP-odd part of a Higgs eigenstate to $WW,ZZ$ is 
one-loop suppressed so that Higgs-strahlung would be dominated by
the CP-even component of the Higgs (or else have a very small 
cross section if the CP-even component is very small).
%As noted earlier, 
Better opportunities are afforded via $t\anti t h$
final state distributions and $\gamma$C and $\mu$C polarization asymmetries
(obtained by varying the polarizations for the colliding $\gamma$'s and $\mu$'s, respectively).

\section{EXOTIC HIGGS SECTORS} 

While a single Higgs doublet is the most economical way to manifest
the Higgs mechanism, simple extensions of the Higgs sector include:
$(i)$ one or more singlet Higgs fields 
[This leads to no particular theoretical problems (or benefits) 
but Higgs discovery can be much more challenging, especially if
there are many singlets],
$(ii)$ more Higgs doublet fields, the simplest
case being the general two-Higgs-doublet model (2HDM)
[In the general 2HDM, CP violation can arise
in the Higgs sector and possibly be responsible for all CP-violating
phenomena],
$(iii)$ (SU(2)$_L$) triplet fields
[In order for $\rho\sim 1$
to be a prediction of the theory (especially to avoid
loop infinities that would require renormalization), 
the vev of any neutral member
of the triplet representation must vanish \cite{Gunion:1991dt}.
Triplets are highly motivated in left-right symmetric models
for the neutrino mass see-saw mechanism. In this case, a right-handed
(SU(2)$_R$) triplet Higgs representation, with
non-zero vev for its neutral member to generate neutrino
masses, requires a partner left-handed (SU(2)$_L$) triplet, 
whose neutral member should have zero vev in order that $\rho=1$  be natural],
$(iv)$ special choices of $T$ and $Y$
for an exotic Higgs multiplet, the next simplest after $T=1/2,|Y|=1$
being $T=3,|Y|=4$, that yield $\rho=1$
at tree level and finite loop corrections to $\rho$ even if
the neutral field has non-zero vev (see \cite{Gunion:1989we}
and references therein).

Coupling constant unification provides further motivation for considering
an extended SM Higgs sector.
For appropriate choices of Higgs representations, it is possible
to achieve coupling constant unification
for SM matter content (i.e. no Susy) \cite{Gunion:1998ii}, although
not at as high a scale as the standard $\mgut\sim 10^{16}\gev$.
For example,  the combination of two 
$T=1/2,Y=1$ and one $T=1,Y=0$ representations gives unification with
$\alpha_s(m_Z)=0.115$ and $\mgut=1.6\times 10^{14}\gev$.
Still lower unification scales, as perhaps appropriate
in theories with extra dimensions, can be achieved for more complicated
Higgs sectors. Thus,
one should not discard complicated Higgs sectors out of hand.

In what follows, we very briefly address the implications for discovery
and precision measurements for some unusual Higgs sectors.
For more theoretical and experimental discussion, see \cite{Gunion:2001vi}.

\subsection{Multi-Singlet Models}

Neither precision electroweak constraints nor LEP2 data
rule out a complicated Higgs sector. In fact,  
%the $1-CL$
%plots from LEP2 as a function of SM-like Higgs boson mass show
%that it is even possible to interpret that data as being due
the LEP2 exclusion plots for a single SM-like Higgs boson -- which
demonstrate a flat $2\sigma$ systematic difference between the expected
and observed background rates -- can be interpreted as indicating
a spread-out Higgs signal, {\it e.g.} several Higgs bosons in the $<114\gev$
region, each with an appropriate fraction of the SM $ZZ$ coupling.
Such a situation was considered in \cite{Espinosa:1998xj}.
The simplest way to achieve this is 
to add a modest number of singlet Higgs fields
to the minimal one-doublet SM Higgs sector. For an appropriate
Higgs potential that mixes the many neutral fields, the
physical Higgs bosons would be mixed states sharing the $WW/ZZ$
coupling strength squared. If these Higgs bosons had masses
spread out every $10-20\gev$ ({\it i.e.}  smaller than the 
detector resolution in a typical decay channel),
a broad/diffuse `continuum' Higgs signal would be the result.

Precision electroweak constraints and also perturbativity for
Higgs sector couplings for all scales between $1\tev$ to $\mgut$ both
imply that the Higgs mass eigenstates with significant $WW,ZZ$
coupling must, on average, have mass below $200-250\gev$.
As shown in \cite{Espinosa:1998xj}, this implies that the broad
diffuse excess in the recoil $M_X$ mass distribution for $\epem\to ZX$
will be observable for $L\gsim 200\fbi$ at a $\rts=500\gev$ LC.
With 1 ab$^{-1}$ of data, it would be possible to map out the $ZZ$
coupling strength as a function of location in $M_X$
and possibly explore branching ratios to various channels as
a function of $M_X$.
In contrast, this broad excess would most likely be very difficult
to detect at the LHC. This would typically be true
even for the $\gamma\gamma$ discovery mode where 
it might be hoped that the excellent mass resolution
would allow observation of a series of narrow peaks. While such
peaks should be carefully searched for even if no other Higgs signal
is seen, they would typically be very suppressed compared
to the SM expectation.  This is because the suppression of the $WW$
coupling to each of the Higgs bosons implies suppression of
the crucial $W$-loop
contribution to the $\gamma\gamma h$ coupling, which would then cancel
substantially against the top-loop contribution.

\subsection{General Two-Higgs-Doublet Models}

In the conventional decoupling limit of a general 2HDM,
there will be a light SM-like Higgs boson which can be
consistent with precision Electroweak constraints and will
be easily detected at both the LHC and the LC.
However, there are non-decoupling scenarios in which the situation
is very different. One such case is that considered in
\cite{Chankowski:2000an} where
the only light Higgs boson has no $WW/ZZ$ couplings
(for example, it could be a light $A$) and all the
other Higgs bosons have mass $\gsim 1\tev$, including a SM-like
CP-even state. The
large negative $\Delta T$ arising from the heavy SM-like
Higgs boson can be compensated by an even larger positive $\Delta T$ 
coming from a small mass splitting between the non-SM-like
heavy Higgs bosons so that the net $S,T$ parameters
fall within the current 90\% precision electroweak ellipse.
The light Higgs boson without $WW/ZZ$ couplings would be very
hard to see at the LHC or a $\rts\sim 800\gev$
LC if $\tanb$ is moderate and its mass is above $\sim 300\gev$.
(For lower masses, $e^+e^-\to \anti\nu \nu$ plus two light Higgs
would allow discovery of the light Higgs.)
The heavy SM-like Higgs boson would be seen at the LHC
but not at the LC until $\rts\gsim 1\tev$ is reached.
Giga-$Z$ constraints on $S,T$ would be
very valuable in fully exploring this type of scenario.
The $\gamma$C and $\mu$C  
would both be able to detect the light, decoupled
Higgs boson over substantial portions of the $\tanb$--Higgs 
mass parameter region
for which it could not otherwise be seen (see \cite{Asner:2001ia} and \cite{Barger:2001mi}, respectively).

\subsection{Extended Higgs sectors in supersymmetric models}

A very attractive extension of the MSSM is the NMSSM (next-to-minimal
supersymmetric model) in which one singlet
Higgs superfield is added to the model \cite{Ellis:1989er}.  The trilinear
superpotential term with the two Higgs doublet superfields and
the singlet superfield yields a natural explanation for the $\mu$
parameter of the MSSM when the scalar component of the Higgs singlet 
superfield acquires a vev.  The prospects for discovering
the three CP-even and two CP-odd Higgs bosons (for purposes
of discussion, we assume CP is conserved) have been explored (see
\cite{Ellwanger:2001iw} and references therein). Discovery of
at least one of the CP-even Higgs bosons would be guaranteed at
a LC with $\rts=500\gev$ and $L\gsim 100\fbi$.
Discovery of one of the CP-even Higgs bosons at the LHC
can also be guaranteed for those regions of NMSSM parameter
space in which heavier Higgs bosons do not decay into lighter Higgs bosons.
Particularly important for this latter conclusion is the recent
development of viable methods for detecting the $t\anti t h(\to b\anti b)$
and $WW\to h\to \tau^+\tau^-$ LHC signals. However, for NMSSM parameter choices
such that there is a light ({\it e.g.} 30 to 50 GeV) pseudoscalar
Higgs boson into which the heavier CP-even Higgs bosons can decay,
there is currently no certainty that even one of the NMSSM
Higgs bosons will be detected at the LHC. A very important
future task will be to develop LHC detection modes that will fill
this void. Still, it is clear that the LC might be absolutely
essential to discover the NMSSM Higgs bosons and would
certainly be required in order to fully explore their properties.

More complicated extensions of the MSSM Higgs sector
are certainly possible, but it is only singlet superfields that can be
easily added without upsetting the nearly perfect coupling constant
unification. Addition of more doublet and or triplet superfields (in pairs,
as required for anomaly cancellation) will destroy coupling
unification unless a carefully chosen set of intermediate-scale
matter superfields are also incorporated in the model.
Coupling constant unification can be retained without intermediate-scale
matter only for certain complicated and carefully chosen 
sets of additional Higgs representations  \cite{Gunion:1998ii}. However,
the unification scale for such models is 
always well below $10^{16}\gev$.  If we stick to the addition
of singlets, the general no-lose theorem of \cite{Espinosa:1998xj}
guarantees that a signal for one of the CP-even Higgs bosons,
or perhaps a spread-out signal from several, will be detectable
at a $\rts=500\gev$ LC. However, for much of the parameter space
of such a model, LHC detection of even one Higgs boson would 
not be possible.

\subsection{Radions from Extra Dimensions}

\begin{figure}[!ht]
\resizebox{7cm}{!}{
\includegraphics[0,150][570,690]{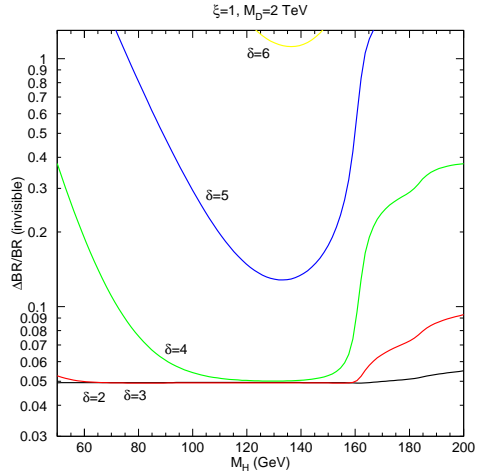}
}
        \caption{Resolution on invisible decays of the Higgs boson in 
the ADD scenario with conformal factor $\xi=1$ and reduced Planck
constant $M_D=2$ TeV.}
        \label{fig:widths1}
\end{figure}
Extra dimensions and related ideas can have a tremendous impact
on Higgs phenomenology.  Most intriguing are the impacts
of the graviscalars present in such theories. These
new scalar degrees of freedom can significantly alter 
Higgs boson phenomenology.   In principle, the graviscalars can
mix with the ordinary Higgs boson through a coupling to the Ricci
scalar.  Most of the relevant phenomenology was addressed
in Ref.~\cite{Giudice:2000av}.
For the ADD scenario, the large number of graviscalar
states can overcome their weak coupling, providing a sizable
invisible width for the Higgs boson.    
Perhaps the best way to 
bound the invisible width is to use the excellent measurement of
the visible width at the LC.  
As demonstrated in Fig.~\ref{fig:widths1},
a good measurement of the BR($h\to$ invisible) is possible for
large $\xi$ and relatively small scales $M_D$ (the invisible width
scales as $\xi^2 m_h^{1+\delta}/M_D^{2+\delta}$).  
This plot is based on applying 
simulations for $m_h=120$ GeV to all Higgs boson masses;
better (worse) resolution is expected for higher (lower) masses
\cite{Schum:2001}.
Even in the limit
$\xi=0$, direct graviscalar production is possible at
colliders (e.g. $e^+e^-\to ZH^{(n)}$), but the effect
is weak and the signal
from spin-2 KK excitations will be substantially larger.
In the RS scenario (with a non-factorizable geometry) there is
only one radion, characterized by its scale $\Lambda_\phi$.
The radion interaction with fermions and electroweak gauge bosons
is similar to the SM one, but scaled by the factor $v/\Lambda_\phi$.
The production of the radion through the typical tree-level processes
is then suppressed by the factor $(v/\Lambda_\phi)^2$.
This compromises
LC searches through the channels $Z^*\to Z\phi,WW\to\phi,b\bar b\phi$,
etc.  
However, the radion has a coupling to $gg$ and
$\gamma\gamma$ pairs 
\noindent through trace anomalies, and the largest 
partial width can be $\Gamma(\phi\to gg)$, thereby dominating
the radion decays, for $m_\phi<2 M_W$.  
Additionally, the
$\xi$ term induces a mixing between the pure radion and Higgs boson.
For $|\xi|\sim .5$ and $v/\Lambda_\phi\sim .1-.2$, observable deviations
from the SM width should be observable at a LC \cite{Hewett:2002rz}.
Typically, the
direct production rate of the radion, even through $gg$ fusion
at the LHC, is small enough to avoid detection.  Surprisingly,
one of the most promising modes at the LHC could be the
observation of $gg\to ZZ\to \ell\ell\ell\ell$ in a rather-narrow
invariant mass range.

\section{SUMMARY AND CONCLUSION}

The analysis of precision Electroweak observables indicates
a high likelihood for the existence of a light SM--like Higgs boson,
which will be discovered possibly 
at the Tevatron and definitely at the LHC.  
While it is possible to obtain agreement with precision
Electroweak data by canceling the large negative $\Delta T$
coming from a heavy SM-like Higgs boson against even
more positive $\Delta T$ contributions from new physics
\cite{Peskin:2001rw,Chankowski:2000an}, 
this is only possible if the SM-like Higgs boson 
has mass below about $1\tev$ (and hence is observable at the LHC)
and if there is other new physics below or at this same scale. 

Whatever the nature of the Higgs sector, it will ultimately be
essential not only to observe all the Higgs bosons but also
to determine their properties at the precision level.
If only a single Higgs boson is observed with decays and production
rates that suggest it is SM-like, it will be crucial to 
search for any deviations of its properties 
from those predicted for the SM Higgs boson.
A complete program will require
the verification that the observed particle carries 
the Higgs boson quantum numbers, and that it is responsible for the 
generation of gauge boson and fermion masses through the model--independent 
measurements of its couplings to the SM particles. 
Deviations from SM predictions could indicate that the SM-like Higgs boson
is part of an extended Higgs sector.  In this case, observation 
of the other Higgs bosons becomes mandatory. In the MSSM, the
other Higgs bosons are most likely to be heavier if the observed
Higgs boson is light and SM-like. In more general models
this need not be the case --- for example, a CP-odd Higgs boson
of moderate mass can easily escape discovery at both the LHC and 
at a LC of modest energy.

A full exploration of the Higgs sector 
can be carried out by a sequence of collider experiments in the
coming decades. While the potential for precise measurements of an observed
Higgs boson at the Tevatron is marginal due to luminosity and energy 
limitations, the LHC will yield a first quantitative picture of the Higgs 
sector, once a significant amount of integrated luminosity 
($\sim$ 200 fb$^{-1}$) is accumulated. Assuming the Higgs is SM-like,
this picture consists of a precise measurement of the Higgs boson mass,
constraints on its spin and quantum numbers, a 
model--dependent determination of its total width, and a measurement
of its coupling to electro-weak gauge bosons at the 5\% level. Access to
the Yukawa couplings is more limited at the LHC. For a light Higgs boson,
only the coupling to the $\tau$ lepton and the top-quark can be measured,
with $\sim$ 10\% precision ($\sim 20\%$ for partial widths).
Beyond that, in a large part of the MSSM parameter space, heavier Higgs bosons
($H, A, H^\pm$) can be observed for large ($\gsim 10$) and small ($\lsim 3$)
values of $\tan\beta$, for masses up to several hundred GeV.

The measurements of the couplings of a SM-like Higgs boson
at the LHC require stringent
assumptions, namely $b/\tau$ universality and absence of important unexpected
decay modes. Lifting these assumptions will increase coupling errors 
substantially. Ultimately, the precision of coupling measurements at 
hadron colliders is systematics limited, by unknown higher order QCD 
corrections to the production cross sections, unknown normalization
of the background, and detector effects.
Some ratios of couplings can be determined more precisely, but these
ratios give incomplete information on the Higgs sector.
If the Higgs sector is more complicated and the hadron colliders
discover more than one Higgs-like signal, an unambiguous
interpretation of these signals will be even more challenging.
Therefore, one concludes that another collider
facility will be necessary to precisely determine the properties
of a SM-like Higgs boson and/or fully delineate the Higgs sector.  

An electron positron linear collider with center--of--mass energy up to 
$\sim$~1~TeV will significantly increase our knowledge about the Higgs sector.
If the lightest Higgs boson is SM-like and has a mass
in the region favored
by EW precision data, already a first stage with $\sqrt{s}\sim 500$~GeV is 
sufficient for a measurement of the essential properties of a Higgs boson
at the percent level. These properties include the mass, quantum numbers and 
couplings to the gauge bosons $Z,W^\pm$ and fermions
$b,c,\tau$. Furthermore, for $m_h \lsim 140$~GeV, a first measurement
of the Higgs trilinear
coupling will be possible, with a relative accuracy of 20\%.
At higher energy ($\sim 1$~TeV) additional important information can be 
obtained from a direct measurement of the top quark Yukawa coupling.
For the specific case of the MSSM,
measurements of these couplings may reveal
distinct differences from the SM that were not accessible at hadron colliders.
However, it is always possible in the MSSM, and many extensions 
of the SM, that the Higgs sector parameters are sufficiently 
into the decoupled  regime that
the lightest Higgs boson is very SM--like.
In this case, LC operations at a higher energy could
discover heavier Higgs bosons for which the LHC 
and a lower energy LC did not provide any evidence.
For example, for a range of values of $\tan\beta$ and $m_A\gsim 300\gev$, the
$H/A$ of the MSSM will not be directly 
visible at the LHC or a low energy LC and the MSSM parameters
can be chosen so that decoupling applies to a very good approximation,
implying that the light $h$ is very SM-like.  Nonetheless, 
the $H/A$ would yield visible signals at a LC with
high energy. (Masses up to $\sim \sqrt{s}/2$ for $e^+e^-$ collisions and up to
$\sim 0.8 \sqrt{s_{ee}}$ for $\gamma\gamma$ collisions
become accessible). 
Even if these heavier Higgs bosons are observable at the LHC,
the LC operating at $\sqrt s \sim 1$ TeV
will provide complementary and generally much more precise measurements
of the couplings of the heavy Higgs bosons.

To summarize, the
proposed $\epem$ linear collider, operating at energies up to
$\sqrt{s}\sim 1$ TeV,
will allow precision measurements of the properties of those Higgs
bosons which are kinematically accessible, which will
almost certainly include a relatively SM-like (or collection
of somewhat SM-like Higgs bosons).
But, there are cases (e.g.~Susy near the decoupling limit)  
in which a complete exploration of the Higgs sector will require
multi--TeV colliders 
(multi-TeV-$e^+e^-$-LC, $\mu$C, VLHC) to complete the
exploration of the Higgs sector through the observation of very heavy
Higgs bosons.

\bibliography{P1WG2}

\end{document}